\documentclass{jpp}
\usepackage{graphicx}

\usepackage[utf8]{inputenc}
\usepackage[T1]{fontenc}
\usepackage{amsmath, amssymb, mathtools}

\usepackage{mathrsfs} % Cursive math font
\usepackage{bm}% bold math
\usepackage{bbold} %mathbb
\usepackage[colorlinks=true,citecolor=blue]{hyperref} % add hypertext capabilities

\newcommand{\pd}[2]{\frac{\partial #1}{\partial #2}}
\newcommand{\dif}{\mathrm{d}}

\newcommand{\grad}{\nabla}

\newcommand{\bra}[1]{\left\langle #1 \right| \!}
\newcommand{\ket}[1]{\! \left| #1 \right\rangle \!}
\newcommand{\braket}[2]{\langle #1 | #2 \rangle}

\newcommand{\im}{\mathsf{i}}

\usepackage{relsize}
\newcommand{\oleft}[1]{\overset{\mathsmaller{\leftarrow}}{#1}}
\newcommand{\oright}[1]{\overset{\mathsmaller{\rightarrow}}{#1}}

\title{Statistical theory of the broadband two-plasmon decay instability}% Force line breaks with \\
\shorttitle{Statistical theory of the broadband two-plasmon decay instability}
\shortauthor{R. T. Ruskov and others}

\author{R. T. Ruskov\aff{1}\corresp{\email{rusko.ruskov@physics.ox.ac.uk}}, R. Bingham\aff{2,3}, L. O. Silva\aff{4,1}, M. Harper\aff{5}, R. Aboushelbaya\aff{1}, J. F. Myatt\aff{6} \and P. A. Norreys\aff{1,7}}
\affiliation{\aff{1}Department of Physics, Atomic and Laser Physics sub-Department, University of Oxford, Clarendon Laboratory, Parks Road, Oxford OX1 3PU, United Kingdom

\aff{2} UKRI-STFC Central Laser Facility, Rutherford Appleton Laboratory, Didcot, Oxon, OX11 0QX, United Kingdom
\aff{3} Department of Physics, University of Strathclyde, John Anderson Building, 107 Rottenrow East, Glasgow, G4 0NG, United Kingdom
\aff{4} GoLP/Instituto de Plasmas e Fus\~ao Nuclear, Instituto Superior T\'ecnico, Universidade de Lisboa, 1049-001 Lisboa, Portugal.
\aff{5} St. Hilda's College, University of Oxford, Cowley Pl, Oxford OX4 1DY, United Kingdom

\aff{6}Department of Electrical and Computer Engineering, University of Alberta, Edmonton, ABT6G 1H9, Canada

\aff{7}John Adams Institute for Accelerator Science, University of Oxford, Denys Wilkinson Building, Keble Road Oxford OX1 3RH, United Kingdom
}

\date{\today}% It is always \today, today,
\begin{document}

\maketitle

\begin{abstract}

There is renewed interest in direct-drive inertial confinement fusion, following the milestone December 2022 3.15 MJ ignition result on the National Ignition Facility. A key obstacle is the control of the two-plasmon decay instability. Here, recent advances in inhomogeneous turbulence theory are applied to the broadband parametric instability problem for the first time. A novel dispersion relation is derived for the two-plasmon decay in a uniform plasma valid under broad-bandwidth laser fields with arbitrary power spectra. The effects of temporal incoherence on the instability are then studied. In the limit of large bandwidth, the well-known scaling relations for the growth rate are recovered, but it is shown that the result is more sensitive to the spectral shape of the laser pulse rather than to its coherence time. The range of wavenumbers of the excited plasma waves is shown to be substantially broadened, suggesting that the absolute instability is favoured in regions further away from the quarter critical density. The intermediate bandwidth regime is explored numerically -- the growth rate is reduced to half its monochromatic value for laser intensities of $10^{15} \, \text{W}/\text{cm}^{2}$ and relatively modest bandwidths of $5 \, \text{THz}$. The instability-quenching properties of a spectrum of discrete lines spread over some bandwidth have also been studied. The reduction in the growth rate is found to be somewhat lower compared to the continuous case but is still significant, despite the fact that, formally, the coherence time of such a laser pulse is infinite. 

\end{abstract}

                              %display desire

%\tableofcontents

\section{Introduction}

Very significant progress has recently been made at the Lawrence Livermore National Laboratory in the generation of high energy density conditions required for fusion energy gain. The August 2021 experiment, where a fusion yield of 1.3 MJ was obtained on the National Ignition Facility, confirmed that the Lawson criterion had been satisfied for the first time \citep{abushwareb2022,zylstra2022}. Subsequently, in December 2022, more energy was released in fusion products than was delivered to target, with exquisite measurements of a 3.12 MJ yield \citep{abu2024achievement}. These results mark a major milestone in the progress of inertial confinement fusion (ICF) research, which have reinvigorated world-wide efforts to explore routes to higher fusion gain for applications such as nuclear stockpile stewardship and inertial fusion energy.  

One of the leading contenders for higher gains at current laser facility energy levels is the direct-drive ICF approach, principally due to the significantly larger energy coupling to the target that results in higher hydrodynamic efficiency of the implosions \citep{craxton2015direct, campbell2021direct}. A major problem for direct-drive ICF is the generation of highly energetic (hot) electrons which can prematurely raise the temperature of the fusion fuel and subsequently degrade the implosion performance. The source of these hot electrons are parametric instabilities that are driven as the laser beams propagate through the coronal plasma surrounding the target. When one of the decay products of a parametric instability is an electron plasma wave -- as is the case for the two-plasmon decay (TPD) instability -- hot electrons are generated \citep{vu2012,vu2012hot}. The fraction of laser energy converted into hot electrons has to be limited to about $0.1 \, \%$ for a successful implosion \citep{craxton2015direct}. 

\begin{figure}
	\includegraphics[width=\textwidth]{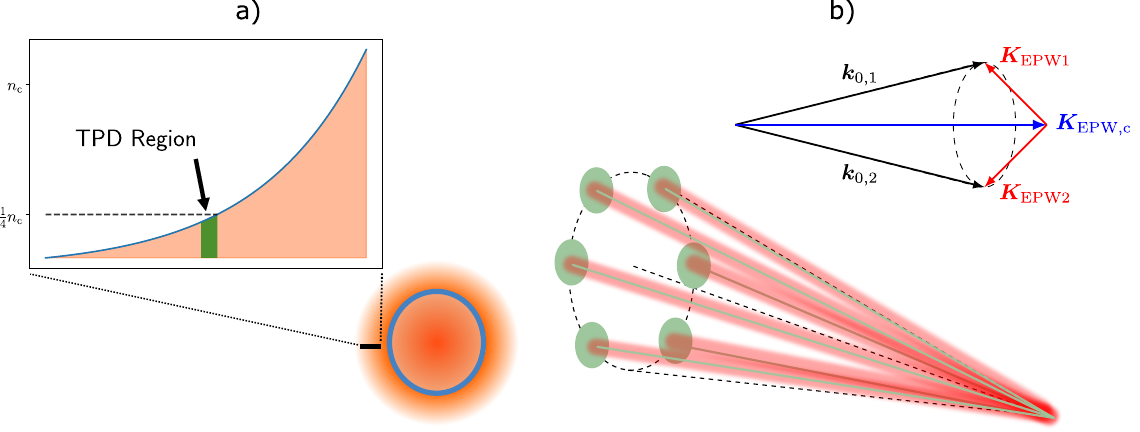}
	\caption{a) An illustration of an imploding ICF pellet, together with its surrounding coronal plasma. The plot shows an illustration of the density gradient in the corona and the region near the quarter critical density where the two-plasmon decay instability is excited. b) An illustration of a set of six laser beams arranged in a cone, as well as a wavevector diagram showing how, due to the cone geometry, a pair of electromagnetic waves ($\bm{k}_{0,1}$ and $\bm{k}_{0,2}$) can share a plasma wave along the cone axis ($\bm{K}_{\text{EPW,c}}$, in blue). The wavenumber matching condition due to the $j^{\text{th}}$ beam in the cone is $\bm{k}_{0,j} = \bm{K}_{\text{EPW,c}} + \bm{K}_{\text{EPW}j}$, where $\bm{K}_{\text{EPW}j}$ (in red) is the second plasma wave involved in TPD which is distinct for each beam.}
	\label{fig:introTPD}
\end{figure}

Progress in the understanding of the impact of the two-plasmon decay on ICF experiments has been considerable in the recent years. TPD is a three-wave instability in which an electromagnetic wave decays into a pair of electron plasma waves. Frequency matching between the waves is possible only in spatially localised region near the quarter critical density (Figure \ref{fig:introTPD}a). Hot electron generation typically occurs when the instability is in the so called \textit{absolute regime} in which wave amplitudes grow exponentially at a fixed point in space; saturation  therefore only occurs through non-linear processes, resulting in a plasma in a highly turbulent state, which accelerates the electrons stochastically \citep{dubois1995saturation,dubois1996saturation,vu2012,vu2012hot}. When multiple laser beams are arranged in a cone (as is the case in real experimental facilities) they can cooperatively drive the instability by sharing a common plasma wave along the axis of symmetry \citep{dubois1992collective} (see Figure \ref{fig:introTPD}b). This results in substantial TPD growth even when the single beam thresholds are far from exceeded \citep{michel2013measured,myatt2014multiple}. While the shared plasma wave along the cone axis is high-$k$ and grows convectively, it has been shown recently that multiple laser beams can drive a low-$k$ absolute mode cooperatively, with a lower threshold compared to the high-$k$ one \citep{zhang2014multiple}. Simulations and experiments of hot electrons driven by the multi-beam TPD instability show that the fraction of laser energy converted into hot electrons is near $0.1 \, \%$ at overlapped intensities of about $10^{15} \, \text{W} \text{cm}^{-2}$, rising to about $1 \, \%$ at $1.5 \times 10^{15} \, \text{W} \text{cm}^{-2}$ \citep{follett2017simulations}. In addition to generating hot electrons, TPD has been recently shown to be responsible for a large amount of anomalous absorption of laser energy (up to $\simeq \! 30 \, \%$) near the quarter critical density region \citep{seka2014nonuniformly,turnbull2020anomalous}. This is likely to be problematic as it increases the distance over which thermal energy needs to be conducted through in order to reach the ablation surface. Taming laser plasma instabilities at high intensities is therefore highly beneficial for direct-drive target designs \citep{paddock2021,paddock2022aa,schmitt2023importance,schmitt2023importance2,trickey2024}. 

\bigskip

It has been known for some time that increasing the temporal bandwidth of the driving laser pulse significantly reduces the growth rates and increases the thresholds for the onset of parametric instabilities \citep{thomson1974,thomson1975finite,laval1976absolute,laval1977,obenschain1976effects,lu1988,lu1989,pesme2007statistical,dubois1992collective}. The Omega Laser Facility at the Laboratory for Laser Energetics is currently undergoing the FLUX upgrade \citep{dorrer2020high,dorrer2021broadband} in which fractional bandwidths of up to $\Delta \omega / \omega_0 = 1\text{-}2\,\%$ will be generated in order to experimentally test these ideas \citep{turnbull2023aps}. Bandwidths of such order are expected to mitigate all laser plasma instabilities in Omega experiments \citep{follett2021thresholds,bates2023}. Here $\Delta \omega$ is the full width at half maximum of the laser power spectrum, and $\omega_0$ the frequency around which it is centred. The thresholds for the onset of parametric instabilities are typically set by inhomogeneity, and calculating them is of great importance. \citet{follett2019} recently compared the analytic model of \citet{lu1989} based on their effective Hamiltonian method against simulations utilising the Laser Plasma Simulation Environment (LPSE) code \citep{myatt2017wave}, and found disagreements as large as a factor of $\sim \! 20 \times$ in the case of the TPD threshold. Given the great promise of controlling laser plasma instabilities with laser bandwidth, it is crucial that theoretical models describe the results of experiments and simulations well, and in some of the most important cases (absolute TPD and Raman scattering) the reliability of current models is somewhat uncertain.   

The broadband parametric instability problem suffers from similar complications as the ones involved in the study of turbulence. Laser bandwidth introduces fluctuations in the laser pulse electric field whose minute details are unknown or too difficult to describe. Hence a statistical approach to the problem must be pursued, only seeking to model some small set of averaged properties of the fluctuations, for example -- their spectra. But difficulties arise as one attempts to carry out statistical predictions about a system's evolution, and one is frequently faced with the statistical closure problem \citep{krommes2002fundamental,krommes2015tutorial,krommes2015statistical}. Most statistical closures such as the commonly used random phase approximation \citep{nazarenko2011wave,connaughton2015rossby} assume homogeneous statistics -- meaning that the statistical properties of the fields of interest do not vary in space. This immediately causes trouble if one is interested in analysing convective growth or non-uniform plasmas -- both of which are crucially important for understanding parametric instabilities \citep{michel2023introduction}. As discussed below it is also problematic even in a uniform plasma, when a zero-order perturbation is present.

Lately there has been some significant progress in moving beyond homogeneous statistical closures. For example, the quasi-linear interaction of inhomogeneous turbulence with plasma has been a longstanding problem where only recently a definitive theory was put forward \citep{dodin2022}. Similarly substantial developments in the theory of drift wave turbulence interacting with zonal flows in tokamaks, have been enabled by new insights about how to properly model the inhomogeneous statistics of the turbulent fields \citep{krommes2015tutorial,krommes2015statistical, parker2016dynamics, ruiz2016zonal,ruiz2019wave,zhu2020,zhu2021wave,parker2014thesis,zhu2020thesis}, improving upon initial work based on the  traditional wave-kinetic approach \citep{smolyakov2000zonal,trines2005quasiparticle,trines2007spontaneous,trines2009applications}. This drift-wave zonal-flow (DW-ZF) problem which is thought important in regulating turbulent transport in tokamaks, also bears resemblance to the broadband parametric instabilities one in the sense that both seek to describe instabilities of some broad spectrum of waves -- the drift wave turbulence on the one hand, and the broad spectrum of electromagnetic waves on the other.

\bigskip

Here a statistical theory of the broadband two-plasmon decay instability in a homogeneous plasma is presented for the first time, valid under laser fields with arbitrary power spectra.  The same closure procedure as that used to study the DW-ZF problem is applied, namely the second order cumulant expansion, also known as CE2, which allows for inhomogeneous statistics. Despite the fact that here one does not impose a density gradient, this is still crucial as inhomogeneity in statistics arises due the strong partially incoherent electromagnetic mode, i.e. the broadband laser pulse which is considered a zero-order perturbation \citep{dubois1976nonlinear,dubois2000kadanoff}. The CE2 closure captures the exact coupling of the stochastic fluctuations to the mean field, and ignores all cumulants of order $\geq 3$\footnote{The first and second order cumulants of a distribution are given by its first and second moments. The cumulants of order $n\geq3$ are the difference between the $n^{\text{th}}$ moment and the value it would have had if the distribution were a Gaussian with the mean and variance implied by the first and second order cumulants. This implies that, for a Gaussian distribution, all cumulants of order $n\geq 3$ vanish.}. This implies that the stochastic fluctuations follow (jointly) normal distributions, and that they interact quasi-linearly with the mean field. The Weyl symbol calculus \citep{mcdonald1988,tracy2014ray,dodin2022} is employed in this article, which greatly facilitates the derivation of the equations governing the correlation functions.  This new work is a generalisation of the approach of \citet{santos2007white}, which considered the broadband Raman scattering instability\footnote{With slight modifications of the same model, the stimulated Brillouin scattering instability has also been studied by \citet{brandao2021bandwidth}.}. Their key insight was that extending the Wigner-Moyal formalism of quantum mechanics provides a natural statistical framework for describing laser pulse propagation inside the plasma, resulting in their Generalised Photon Kinetics (GPK) model \citep{santos2005wigner,silva2013theory}. GPK is in essence, the Wigner-Moyal representation of the Klein-Gordon equation which describes laser pulse propagation in a plasma. They did not face a closure problem as they assumed the electrostatic fluctuations associated with electron plasma waves to be non-random. In the case of the two-plasmon decay this is not possible, and one has to provide a statistical description of the plasma as well. It should be noted that were they to allow for stochasticity of the plasma, then they would effectively use the same closure as the one described here, and their model would have the same statistical content as the new one presented here. 

The paper is organised as follows. In Section \ref{sec:model} the dispersion relation for the two-plasmon decay instability is derived, one that is valid for laser fields with arbitrary power spectra in a uniform plasma. Emphasis is made on the details of the statistical closure. In Section \ref{sec:applications} the broadband dispersion relation is applied to some cases of interest -- namely, it is shown that it reduces to the well known dispersion relation for a single monochromatic beam, as well as multiple monochromatic non-interfering ones; following that, the effects of temporal incoherence on the instability as driven by a single broadband laser beam are considered; and lastly the extent to which a laser power spectrum consisting of multiple discrete spectral lines spread over a bandwidth $\Delta \omega$ approximates the instability-quenching properties of a continuous spectrum is explored. To conclude the article, a summary and a discussion of the new results is presented in Section \ref{sec:summary}.

\section{Model}
\label{sec:model}

The two-plasmon decay instability is three-wave parametric instability in which an electromagnetic wave decays into a pair of plasma waves. Due to the Manley-Rowe relations, the frequencies and wavenumbers of the interacting waves must be matched \citep{michel2023introduction}. This localises the instability spatially near the quarter critical density -- the point where the laser frequency is equal to twice the local plasma frequency (see Figure \ref{fig:introTPD}a). Furthermore, since all of the waves are high in frequency, the ion motion can be neglected in the linear stage of the instability. The response of the electrons in the presence of the laser field can be analysed using non-relativistic fluid equations. Kinetic effects can be ignored  provided $k \lambda_{\text{D}e} \ll 1$. Relativistic effects are negligible under conventional hot-spot ignition ICF where the laser intensity is of order $I_{\text{L}} \sim 10^{15} \, \text{W}/\text{cm}^{2}$ and the laser wavelength is $\lambda_{\text{L}} \sim 350 \, \mu \text{m}$. This is because the quiver velocity of the electrons $\boldsymbol{v}_{\text{os}}(t,\boldsymbol{x}) = e \boldsymbol{A}_0/m_e$ in the presence of an electromagentic wave described by the vector potential $\boldsymbol{A}_0(t,\boldsymbol{x})$ is $v_{\text{os}}/c \sim 0.01$ under such conditions. Finally, for the motion of the ions to be neglected, we need the amplitude of electrostatic waves to be low enough such that $\sqrt{W_e} \ll k \lambda_{\text{D}e}$, where $W_e = \tfrac12 \epsilon_0 |\bm{\grad} \varphi|^2/n_e T_e$. This ensures that the nonlinear terms which couple the electron and ion motion are higher order and negligible. So, the following ordering will be assumed from now on:
	\begin{equation}
		\sqrt{\frac{\epsilon_0 |\bm{\grad} \varphi|^2}{2n_e T_e}} \ll \frac{v_{\text{os}}}{c} \sim k \lambda_{\text{D}e} \ll 1.
	\end{equation}
	
Assuming a purely electrostatic perturbation, linearising the continuity and momentum equations for the electron fluid gives \citep{liu1976,simon1983inhomogeneous}:
\begin{equation}\label{tpdeqs}
  	\begin{aligned}
  		\pd{n}{t} &= - \grad^2 \psi - \boldsymbol{v}_{\text{os}} \boldsymbol{\cdot} \boldsymbol{\grad} n, \\
  		\pd{\psi}{t} &= \omega_{\text{pe}}^{2} \grad^{-2} n - 3 v_{\text{th}e}^{2} n - \boldsymbol{v}_{\text{os}} \boldsymbol{\cdot} \boldsymbol{\grad} \psi.
  	\end{aligned}
\end{equation}
Here $n(t,\boldsymbol{x})$ is the electron density perturbation normalised to the equilibrium density $n_{0}$ which will be assumed uniform for simplicity, $\psi(t,\boldsymbol{x})$ is the velocity potential of the electrons meaning $\boldsymbol{u} = \grad \psi$ where $\boldsymbol{u}$ is the electron fluid velocity, $\omega_{\text{pe}}^{2} = e^2 n_0/\epsilon_0 m_e$ is the equilibrium plasma frequency, $v_{\text{th}e}^{2} = T_e/m_e$ is the electron thermal velocity. Since the goal is to develop a description of the instability under broad-bandwidth (stochastic) laser fields, $n$, $\psi$ and $\bm{v}_{\text{os}}$ are assumed to be random variables.

The wave equation governing the electron plasma waves is:
\begin{equation}
\label{waveeq}
	\left( \pd{^2}{t^2} + \omega_{\text{pe}}^{2} - 3 v_{\text{th}e}^2 \grad^2 \right) \! n = \grad^2 \left( \boldsymbol{v}_{\text{os}} \boldsymbol{\cdot} \boldsymbol{\grad} \psi \right) - \pd{}{t} \left( \boldsymbol{v}_{\text{os}} \boldsymbol{\cdot} \boldsymbol{\grad} n \right),
\end{equation}
where the right side contains the TPD driving terms. One can express the driving terms through the Wigner functions describing the correlations between the pump field $\boldsymbol{v}_{\text{os}}$ and the plasma waves, as has been done in the case of inhomogeneous Navier-Stokes turbulence \citep{tsiolis2020structure} (see Appendix \ref{appDriving}):
\begin{equation}\label{eqDrivTerms}
	\boldsymbol{v}_{\text{os}} \boldsymbol{\cdot} \boldsymbol{\grad} \psi = \im \int \frac{\dif \boldsymbol{k}}{(2\pi)^3} \, \boldsymbol{k}^{\top} \star \bm{\mathsf{W}}_{\psi \boldsymbol{v}}^{\top}.
\end{equation}
Here $\bm{\mathsf{W}}_{\psi \boldsymbol{v}}(\boldsymbol{x},\boldsymbol{k}) = \mathscr{W} \! \left[ \, \ket{\psi} \bra{\boldsymbol{v}_{\text{os}}} \, \right]$, with $\mathscr{W}$ being the Wigner transform, and the bra-ket notation is used as in \citet{dodin2022}; $\bm{\mathsf{W}}_{\psi \boldsymbol{v}}^{\top}$ denotes the transpose of $\bm{\mathsf{W}}_{\psi \boldsymbol{v}}$. The Wigner transform $\mathscr{W}[\hat{A}]$ of some generic operator $\hat{A}$ is called the Weyl symbol of $\hat{A}$, and is defined as:
\begin{equation*}
	\mathsf{A}(\boldsymbol{x},\boldsymbol{k}) \equiv \mathscr{W}[\hat{A}] \doteq \int \dif \boldsymbol{s} \, e^{- \im \boldsymbol{k} \boldsymbol{\cdot} \boldsymbol{s}} \left \langle \boldsymbol{x} + \frac{\boldsymbol{s}}{2} \right | \hat{A} \left | \boldsymbol{x} - \frac{\boldsymbol{s}}{2} \right \rangle.
\end{equation*}
The Moyal star product $\star$ between two symbols is defined as $\mathsf{A} \star \mathsf{B} \doteq \mathscr{W}[\hat{A} \hat{B}]$; it is given by: $\mathsf{A}(\boldsymbol{x},\boldsymbol{k}) \star \mathsf{B}(\boldsymbol{x},\boldsymbol{k}) =  \mathsf{A}(\boldsymbol{x},\boldsymbol{k}) e^{\im \hat{\mathcal{P}}/2}  \mathsf{B}(\boldsymbol{x},\boldsymbol{k})$, with $\hat{\mathcal{P}}$ being the Poisson bracket $\hat{\mathcal{P}} = \oleft{\partial_{\boldsymbol{x}}} \boldsymbol{\cdot} \oright{\partial_{\boldsymbol{k}}} - \oleft{\partial_{\boldsymbol{k}}} \boldsymbol{\cdot} \oright{\partial_{\boldsymbol{x}}}$, and the arrows indicating the directions in which the derivatives are acting. Element-wise application of the Moyal star product $\star$ and the usual matrix multiplication rules are assumed, meaning that for any two matrix symbols $\bm{\mathsf{A}}$ and $\bm{\mathsf{B}}$ one applies the Moyal product as follows: $\left( \bm{\mathsf{A}} \star \bm{\mathsf{B}} \right)_{ij} = \sum_k (\bm{\mathsf{A}})_{ik} \star (\bm{\mathsf{B}})_{kj}$. A brief introduction to the Weyl symbol calculus can be found in Appendix \ref{appWeyl}. With these definitions, one can see that the quantity $\bm{\mathsf{W}}_{\psi \boldsymbol{v}}(\boldsymbol{x},\boldsymbol{k})$ is given by:
\begin{equation}
	\bm{\mathsf{W}}_{\psi \boldsymbol{v}}(\boldsymbol{x},\boldsymbol{k}) = \int \dif \boldsymbol{s} \, e^{- \im \boldsymbol{k} \boldsymbol{\cdot} \boldsymbol{s}} \, \psi( \boldsymbol{x} + \boldsymbol{s}/2 ) \boldsymbol{v}_{\text{os}}^{\dagger} (\boldsymbol{x} - \boldsymbol{s}/2).
\end{equation}
After ensemble averaging $\bm{\mathsf{W}}_{\psi \boldsymbol{v}}(\boldsymbol{x},\boldsymbol{k})$ will represent the correlations of the laser pulse field $\boldsymbol{v}_{\text{os}}$ with the the velocity potential $\psi$. The second driving term is rewritten in a completely analogous manner leading to the quantity $\bm{\mathsf{W}}_{n \boldsymbol{v}}(\boldsymbol{x},\boldsymbol{k})$ which will represent the correlations between $\boldsymbol{v}_{\text{os}}$ and the density perturbation $n$. 
 
Taking the Fourier transform of the wave equation, and ensemble averaging leads to:
\begin{equation}\label{disp}
  	D_e(\Omega, \boldsymbol{K}) \mathring{\overline{n}} (\Omega,\boldsymbol{K}) = \int \frac{\dif \boldsymbol{k}}{(2\pi)^3} \, \left( \boldsymbol{k} - \tfrac12 \boldsymbol{K} \right) \boldsymbol{\cdot} \left( \im K^2 \mathring{\overline{\bm{\mathsf{W}}}}_{\psi \boldsymbol{v}} - \Omega \mathring{\overline{\bm{\mathsf{W}}}}_{n \boldsymbol{v}} \right), 
\end{equation}
where $D_e(\Omega, \boldsymbol{K}) = \Omega^2 - \omega_{\text{pe}}^{2} - 3 v_{\text{th}e}^{2} K^2 = \Omega^2 - \omega_{e\boldsymbol{K}}^2$ is the dispersion function for the electron plasma waves, and the Fourier transformed quantities are denoted as follows: $\mathring{n}(\Omega,\boldsymbol{K}) = \int \dif t \dif \boldsymbol{x}\, n(t,\boldsymbol{x}) e^{\im \Omega t - \im \boldsymbol{K} \boldsymbol{\cdot} \boldsymbol{x}}$. The over-line represents ensemble averaging. Here one now faces the closure problem -- to calculate the mean field $\overline{n}$,  knowledge is needed of second order quantities: $\overline{\bm{\mathsf{W}}}_{\psi \boldsymbol{v}}$ and $\overline{\bm{\mathsf{W}}}_{n \boldsymbol{v}}$. The powerful toolbox of the Weyl symbol calculus is now applied to derive and solve the equations governing these quantities.

The basic idea is as follows: if one writes down the governing equations of the system in Schr\"odinger form, then all of the second order correlation functions describing the system will be governed by the Wigner-Moyal equation (see Appendix \ref{appWeyl}). A similar procedure was carried out for the case of inhomogeneous fluid turbulence by \citet{tsiolis2020structure}. The TPD equations (\ref{tpdeqs}) are first order in time and are readily put in Schr\"odinger form. The vector potential $\boldsymbol{A}_{0}$ describing the propagation of an electromagnetic mode in a plasma obeys a Klein-Gordon equation and therefore $\left( \partial_t^2 - c^2 \grad^2 + \omega_{\text{pe}}^{2} \right) \! \boldsymbol{v}_{\text{os}} = 0$, since $\boldsymbol{v}_{\text{os}}$ differs from $\boldsymbol{A}_0$ only by a constant factor \citep{kruer2019physics,michel2023introduction}. There are no source or sink terms in this equation since the linear stage of the instability is considered and hence pump depletion is ignored. This second order in time equation can be decomposed into two first order ones by defining the auxiliary fields $\bm{\phi}, \bm{\chi} = \frac12 \! \left( \boldsymbol{v}_{\text{os}} \pm \im \omega_{\text{pe}}^{-1} \partial_t \boldsymbol{v}_{\text{os}} \right)$ \citep{santos2005wigner}. 

The system of the decomposed Klein-Gordon equation, together with the two TPD equations are then written as a Schr\"odinger equation: $\im \partial_t \bm{\Psi} = \bm{H} \bm{\Psi}$, with the following matrix Hamiltonian and state vector:
\begin{equation}\label{eqHW}
	\bm{H} = \begin{pmatrix}
		0 & - \im \grad^2 & - \im \bm{\eta}^{\top} & - \im \bm{\eta}^{\top}\\
		\im \left( \omega_{\text{pe}}^{2} \grad^{-2} - 3 v_{\text{th}e}^{2} \right) & 0 & - \im \boldsymbol{u}^{\top} & - \im \boldsymbol{u}^{\top} \\
		0 & 0 & - \frac{c^2}{2 \omega_{\text{pe}}} \grad^2  + \omega_{\text{pe}} & - \frac{c^2}{2 \omega_{\text{pe}}} \grad^2 \\
		0 & 0 & \frac{c^2}{2 \omega_{\text{pe}}} \grad^2 &  \frac{c^2}{2 \omega_{\text{pe}}} \grad^2 - \omega_{\text{pe}}
	\end{pmatrix}, \quad  \bm{\Psi} = \begin{pmatrix}
		n \\ \psi \\ \bm{\phi} \\ \bm{\chi}
	\end{pmatrix},
\end{equation}
where $\bm{\eta} \doteq \grad n$ and $\boldsymbol{u} \doteq \grad \psi$. The Wigner matrix $\bm{\mathsf{W}} = \mathscr{W} \! \left[ \, \ket{\bm{\Psi}} \bra{\bm{\Psi}} \,  \right]$ satisfies the Wigner-Moyal equation:
\begin{equation}
	\im \partial_t \bm{\mathsf{W}} = \bm{\mathsf{H}} \star \bm{\mathsf{W}} - \bm{\mathsf{W}} \star \bm{\mathsf{H}}^{\dagger},
\end{equation}
with $\bm{\mathsf{H}} \doteq \mathscr{W}[\bm{H}]$ being the symbol of the matrix Hamiltonian. It is useful to write out the contents of the Wigner matrix and the symbol of the matrix Hamiltonian explicitly:
\begin{align}
  	 \bm{\mathsf{H}} =& \begin{pmatrix}
		0 & \im k^2  & - \im \bm{\eta}^{\top} & - \im \bm{\eta}^{\top}\\
		- \im k^{-2} \omega_{e\boldsymbol{k}}^2 & 0 & - \im \boldsymbol{u}^{\top} & - \im \boldsymbol{u}^{\top}\\
		0 & 0 &  \frac{c^2}{2 \omega_{\text{pe}}} k^2  + \omega_{\text{pe}} & \frac{c^2}{2 \omega_{\text{pe}}} k^2 \\
		0 & 0 & - \frac{c^2}{2 \omega_{\text{pe}}} k^2 & - \frac{c^2}{2 \omega_{\text{pe}}} k^2 - \omega_{\text{pe}}
	\end{pmatrix}, \\
	\bm{\mathsf{W}} =& \begin{pmatrix}
  		\mathsf{W}_{nn} & \mathsf{W}_{n \psi} & \bm{\mathsf{W}}_{n \bm{\phi}} & \bm{\mathsf{W}}_{n \bm{\chi}} \\
  		\mathsf{W}_{\psi n} & \mathsf{W}_{\psi \psi } & \bm{\mathsf{W}}_{\psi \bm{\phi}} & \bm{\mathsf{W}}_{\psi \bm{\chi}} \\
  		\bm{\mathsf{W}}_{\bm{\phi} n} & \bm{\mathsf{W}}_{\bm{\phi} \psi} & \bm{\mathsf{W}}_{\bm{\phi} \bm{\phi}} & \bm{\mathsf{W}}_{\bm{\phi} \bm{\chi}} \\
  		\bm{\mathsf{W}}_{\bm{\chi} n} & \bm{\mathsf{W}}_{\bm{\chi} \psi} & \bm{\mathsf{W}}_{\bm{\chi} \bm{\phi}} & \bm{\mathsf{W}}_{\bm{\chi} \bm{\chi}}
  	\end{pmatrix},
\end{align}
where the quantities which have a vector index in the second slot such as $\bm{\mathsf{W}}_{n \bm{\phi}}$ are row vectors; similarly $\bm{\mathsf{W}}_{\bm{\phi}n}$ is a column vector; and quantities such as $\bm{\mathsf{W}}_{\bm{\phi} \bm{\phi}}$ with two vector indices are matrices.

Note that the relevant Wigner functions in (\ref{disp}) are related to the components of this Wigner matrix above through $\bm{\mathsf{W}}_{n \boldsymbol{v}} = \bm{\mathsf{W}}_{n \bm{\phi}} + \bm{\mathsf{W}}_{n \bm{\chi}}$ and $\bm{\mathsf{W}}_{\psi \boldsymbol{v}} = \bm{\mathsf{W}}_{\psi \bm{\phi}} + \bm{\mathsf{W}}_{\psi \bm{\chi}}$, and so one needs only to compute the evolution equations of the upper right $2\times 2$ corner of $\bm{\mathsf{W}}$. Doing so and taking sums and differences of the equations in the resulting system, one gets equations for $\bm{\mathsf{W}}_{n \boldsymbol{v}}$ and $\bm{\mathsf{W}}_{\psi \boldsymbol{v}}$, which upon ensemble averaging results in the following system:
\begin{equation}\label{system-noFT}
  	\begin{aligned}
  		\im \partial_t \overline{\bm{\mathsf{W}}}_{n \boldsymbol{v}} &= \im k^2 \star \overline{\bm{\mathsf{W}}}_{\psi \boldsymbol{v}} - \im \overline{\bm{\eta}}^{\top} \star \overline{\bm{\mathsf{W}}}_{\boldsymbol{v} \boldsymbol{v}} - \omega_{\text{pe}} \overline{\bm{\mathsf{B}}}_{n \boldsymbol{v}}, \\
  		\im \partial_t \overline{\bm{\mathsf{B}}}_{n \boldsymbol{v}} &= \im k^2 \star \overline{\bm{\mathsf{B}}}_{\psi \boldsymbol{v}} - \im \overline{\bm{\eta}}^{\top} \star \overline{\bm{\mathsf{B}}}_{\boldsymbol{v} \boldsymbol{v}} - \omega_{\text{pe}}^{-1} \overline{\bm{\mathsf{W}}}_{n \boldsymbol{v}} \star \omega_{\boldsymbol{k}}^2, \\
  		\im \partial_t \overline{\bm{\mathsf{W}}}_{\psi \boldsymbol{v}} &= - \im k^{-2} \omega_{e\boldsymbol{k}}^2 \star \overline{\bm{\mathsf{W}}}_{n \boldsymbol{v}} - \im \overline{\boldsymbol{u}}^{\top} \star \overline{\bm{\mathsf{W}}}_{\boldsymbol{v} \boldsymbol{v}} - \omega_{\text{pe}} \overline{\bm{\mathsf{B}}}_{\psi \boldsymbol{v}}, \\
  		\im \partial_t \overline{\bm{\mathsf{B}}}_{\psi \boldsymbol{v}} &= - \im k^{-2} \omega_{e\boldsymbol{k}}^2 \star \overline{\bm{\mathsf{B}}}_{n \boldsymbol{v}} - \im \overline{\boldsymbol{u}}^{\top} \star \overline{\bm{\mathsf{B}}}_{\boldsymbol{v} \boldsymbol{v}} - \omega_{\text{pe}}^{-1} \overline{\bm{\mathsf{W}}}_{\psi \boldsymbol{v}} \star \omega_{\boldsymbol{k}}^2.
  	\end{aligned}
\end{equation}
Here $\omega_{\boldsymbol{k}}^2 = \omega_{\text{pe}}^{2} + k^2 c^2$ is the electromagnetic wave frequency. All of the $\bm{\mathsf{B}}$ quantities are not of interest and simply facilitate the calculation; they are defined as follows: $\bm{\mathsf{B}}_{n \boldsymbol{v}} \doteq \bm{\mathsf{W}}_{n \bm{\phi}} - \bm{\mathsf{W}}_{n \bm{\chi}}$, $\bm{\mathsf{B}}_{\psi \boldsymbol{v}} \doteq \bm{\mathsf{W}}_{\psi \bm{\phi}} - \bm{\mathsf{W}}_{\psi \bm{\chi}}$, $\bm{\mathsf{B}}_{\boldsymbol{v} \boldsymbol{v}} \doteq \bm{\mathsf{W}}_{\bm{\phi} \bm{\phi}} + \bm{\mathsf{W}}_{\bm{\chi} \bm{\phi}} - \bm{\mathsf{W}}_{\bm{\phi} \bm{\chi}} - \bm{\mathsf{W}}_{\bm{\chi} \bm{\chi}}$. Here, the fact that $\bm{\mathsf{W}}_{\boldsymbol{v} \boldsymbol{v}} = \bm{\mathsf{W}}_{\bm{\phi} \bm{\phi}} + \bm{\mathsf{W}}_{\bm{\chi} \bm{\phi}} + \bm{\mathsf{W}}_{\bm{\phi} \bm{\chi}} + \bm{\mathsf{W}}_{\bm{\chi} \bm{\chi}}$ is used, which follows from the linearity of the Wigner transform.

The crucial step which is the essence of the CE2 closure, is the splitting of the average of the product, into the product of the averages, for the third-order quantities: $\overline{\bm{\eta}^{\top} \star \bm{\mathsf{W}}} = \overline{\bm{\eta}}^{\top} \star \overline{\bm{\mathsf{W}}}$. This effectively assumes the third-order cumulant $\overline{ \tilde{\bm{\eta}}^{\top} \star \tilde{\bm{\mathsf{W}}}}$ to be negligible (the tilde denotes the fluctuating part of the quantity) \citep{krommes2015statistical}. It is exactly zero for fields obeying jointly normal (Gaussian multivariate) statistics. In contrast to turbulence, here this assumption is easier to justify. One does not expect the interactions between the fluctuations to be of interest, which is what is being neglected when one discards higher order cumulants and assumes Gaussianity. They wouldn't be of interest since such effects are what leads to fluctuation intermittency, or are responsible for the typical turbulent energy cascades, neither of which is assumed here to be of relevance in the linear stage of the instability. The fluctuations in the electric field of the laser pulse are prescribed in a sense (their power spectrum $\overline{\bm{\mathsf{W}}}_{\boldsymbol{v} \boldsymbol{v}}$ is given), and are expected to be normal for both temporally incoherent pulses \citep{goodman2015statistical}, as well as transversely incoherent fields produced by a random phase plate for example  \citep{rose1993statistical,garnier1999statistics}, where in both cases this is a consequence of the central limit theorem. Hence the driving field $\boldsymbol{v}_{\text{os}}$ is safely assumed a Gaussian random process. It should be noted that the Wigner-Moyal formalism is also very useful in deriving more sophisticated closures such as the quasi-normal one as shown by \citet{ruiz2019wave}.

The Wigner function is bilinear and therefore $\bm{\mathsf{B}}_{\boldsymbol{v} \boldsymbol{v}} = \bm{\mathsf{W}}_{(\bm{\phi} + \bm{\chi)}(\bm{\phi} - \bm{\chi)}}$. If the laser field $\boldsymbol{v}_{\text{os}}$ is statistically stationary, it is shown in Appendix \ref{appProofPump} that the ensemble averaged Wigner function $\overline{\bm{\mathsf{W}}}_{\boldsymbol{v}\boldsymbol{v}}$ and the associated quantity $\overline{\bm{\mathsf{B}}}_{\boldsymbol{v} \boldsymbol{v}}$ are related as follows: $\overline{\bm{\mathsf{B}}}_{\boldsymbol{v} \boldsymbol{v}}(\boldsymbol{k}) = \omega_{\text{pe}}^{-1} \omega_{\boldsymbol{k}} \overline{\bm{\mathsf{W}}}_{\boldsymbol{v}\boldsymbol{v}}(\boldsymbol{k})$. This is an important step in the derivation as it makes all symbol products in (\ref{system-noFT}) have one of the quantities being a function of $\boldsymbol{x}$ or $\boldsymbol{k}$ only. Taking the Fourier transform of the system results in:
\begin{equation}\label{system}
  	\begin{aligned}
  		-\Omega \mathring{\overline{\bm{\mathsf{W}}}}_{n \boldsymbol{v}} &= \im k^2_- \mathring{\overline{\bm{\mathsf{W}}}}_{\psi \boldsymbol{v}} - \im \mathring{\overline{\bm{\eta}}}^{\top} \overline{\bm{\mathsf{W}}}_{\boldsymbol{v} \boldsymbol{v}}^{+} - \omega_{\text{pe}} \mathring{\overline{\bm{\mathsf{B}}}}_{n \boldsymbol{v}}, \\
  		-\Omega \mathring{\overline{\bm{\mathsf{B}}}}_{n \boldsymbol{v}} &= \im k_-^2  \mathring{\overline{\bm{\mathsf{B}}}}_{\psi \boldsymbol{v}} - \im \omega_{\text{pe}}^{-1} \omega_{\boldsymbol{k}_+} \mathring{\overline{\bm{\eta}}}^{\top} \overline{\bm{\mathsf{W}}}_{\boldsymbol{v} \boldsymbol{v}}^{+}  - \omega_{\text{pe}}^{-1} \omega_{\boldsymbol{k}_+}^2 \mathring{\overline{\bm{\mathsf{W}}}}_{n \boldsymbol{v}}, \\
  		-\Omega \mathring{\overline{\bm{\mathsf{W}}}}_{\psi \boldsymbol{v}} &= - \im k^{-2}_- \omega_{e\boldsymbol{k}_-}^2 \mathring{\overline{\bm{\mathsf{W}}}}_{n \boldsymbol{v}} - \im \mathring{\overline{\boldsymbol{u}}}^{\top} \overline{\bm{\mathsf{W}}}_{\boldsymbol{v} \boldsymbol{v}}^{+} - \omega_{\text{pe}} \mathring{\overline{\bm{\mathsf{B}}}}_{\psi \boldsymbol{v}}, \\
  		-\Omega \mathring{\overline{\bm{\mathsf{B}}}}_{\psi \boldsymbol{v}} &= - \im k^{-2}_- \omega_{e\boldsymbol{k}_-}^2 \mathring{\overline{\bm{\mathsf{B}}}}_{n \boldsymbol{v}} - \im \omega_{\text{pe}}^{-1} \omega_{\boldsymbol{k}_+} \mathring{\overline{\boldsymbol{u}}}^{\top} \overline{\bm{\mathsf{W}}}_{\boldsymbol{v} \boldsymbol{v}}^{+} -\omega_{\text{pe}}^{-1} \omega_{\boldsymbol{k}_+}^2  \mathring{\overline{\bm{\mathsf{W}}}}_{\psi \boldsymbol{v}},
  	\end{aligned}
\end{equation}	
where the quantities with superscripts $\pm$ are evaluated at $\boldsymbol{k}_{\pm} = \boldsymbol{k} \pm \tfrac12 \boldsymbol{K}$, $\mathring{\overline{\bm{\eta}}}$ and $\mathring{\overline{\boldsymbol{u}}}$ are functions of $\boldsymbol{K}$ only (the Fourier image of $\boldsymbol{x}$), and all other quantities are functions of both $\boldsymbol{k}$ and $\boldsymbol{K}$. 

%The standard derivation of TPD makes an important approximation which is equivalent to ignoring terms of order $(v_{\text{os}}/c)^2 \ll 1$ in the resulting expression for the TPD growth rate \citep{kruer2019physics}.

In order to keep the model self-consistent under the assumed ordering, the growth rate of the instability can be calculated only up to order $v_{\text{os}}/c$ and higher order terms need to be discarded \citep{kruer2019physics}. The way this works in the present case is as follows. To close the system (\ref{system}), $\mathring{\boldsymbol{u}}$ can be expressed in terms of $\mathring{\bm{\eta}}$ and $\mathring{\bm{\mathsf{W}}}_{\psi \boldsymbol{v}}$ through the second equation in the TPD system (\ref{tpdeqs}). Then the equations containing $\mathring{\boldsymbol{u}}$ will involve terms which are $\propto \mathring{\bm{\mathsf{W}}}_{\psi \boldsymbol{v}} \mathring{\bm{\mathsf{W}}}_{\boldsymbol{v} \boldsymbol{v}}$ As in the standard derivation in \citet{kruer2019physics}, this introduces terms of order $(v_{\text{os}}/c)^2 \ll 1$ in the growth rate which are ignored, and therefore in (\ref{system}) one uses the approximation: $ \mathring{\boldsymbol{u}} = \im (\Omega/K^2) \mathring{\bm{\eta}}$. 

Solving the system for the quantities of interest: $\mathring{\overline{\bm{\mathsf{W}}}}_{\psi \boldsymbol{v}}$ and $\mathring{\overline{\bm{\mathsf{W}}}}_{n \boldsymbol{v}}$, in terms of the laser field $\overline{\bm{\mathsf{W}}}_{\boldsymbol{v} \boldsymbol{v}}$ and substituting them into (\ref{disp}), results in the homogeneous broadband TPD dispersion relation:
%\begin{widetext}
\begin{equation}\label{broadTPD}
	D_e(\Omega,\boldsymbol{K}) = \int \! \frac{\dif \boldsymbol{k}}{(2\pi)^3} \frac{\left( k_-^2 \Omega - K^2 \Omega_- \right)^2 }{K^2 k_-^2 \, D_e(\Omega_-,\boldsymbol{k}_-)}  \,  \boldsymbol{k}_- \boldsymbol{K}^{\top} \! \boldsymbol{:} \!\overline{\bm{\mathsf{W}}}_{\boldsymbol{v} \boldsymbol{v}} (\boldsymbol{k}),
\end{equation}
%\end{widetext}
where $\Omega_- = \omega_{\boldsymbol{k}} - \Omega$, the integral is to be taken along the Landau contour, and $\boldsymbol{k}_- \doteq \boldsymbol{k} - \boldsymbol{K}$ has been redefined, and the colon denotes double matrix contraction\footnote{The double matrix contraction is defined as $\boldsymbol{A} \boldsymbol{:} \boldsymbol{B} = \sum_{ij} A_{ij} B_{ji}$. For matrices which can be written as $\boldsymbol{A} = \boldsymbol{a}_1\boldsymbol{a}_2^{\top}$ and $\boldsymbol{B} = \boldsymbol{b}_1 \boldsymbol{b}_2^{\top}$ one has $\boldsymbol{A} \boldsymbol{:} \boldsymbol{B} = (\boldsymbol{a}_1 \boldsymbol{\cdot} \boldsymbol{b}_2) (\boldsymbol{a}_2 \boldsymbol{\cdot} \boldsymbol{b}_1)$.}. This is the main result of the new model. For a statistically stationary random process, the two point correlation function and the power spectrum form a Fourier pair (Wiener-Khinchin theorem) \citep{goodman2015statistical}. So one can notice that the expression has a familiar form and a simple physical interpretation: the broadband dispersion relation is the monochromatic one integrated over the laser field power spectrum $\overline{\bm{\mathsf{W}}}_{\boldsymbol{v} \boldsymbol{v}} (\boldsymbol{k})$:
\begin{equation}
	\overline{\bm{\mathsf{W}}}_{\boldsymbol{v} \boldsymbol{v}} (\boldsymbol{k}) = \int \dif \boldsymbol{s} \, e^{- \im \boldsymbol{k} \boldsymbol{\cdot} \boldsymbol{s}} \, \overline{\boldsymbol{v}_{\text{os}}(\boldsymbol{x}+\boldsymbol{s}/2) \boldsymbol{v}^\dagger_{\text{os}}(\boldsymbol{x}-\boldsymbol{s}/2)}.
\end{equation} 
This dispersion relation allows one to study the homogeneous two-plasmon decay under laser fields with arbitrary power spectra.

\section{Applications}
\label{sec:applications}

\subsection{Monochromatic beams}

The broadband TPD dispersion relation (\ref{broadTPD}) reduces to the well known  result for a monochromatic plane wave. Consider a pump of the form $\boldsymbol{v}_{\text{os}}(t,\boldsymbol{x}) = \tfrac12 v_{\text{os}} \bm{\sigma} e^{\im \boldsymbol{k}_0 \boldsymbol{\cdot} \boldsymbol{x} - \im \omega_0t} + \text{c.c.}$, where $\omega_0 \equiv \omega_{\boldsymbol{k}_0}$ is the laser frequency, and $\bm{\sigma}$ is the laser polarisation. $\bm{\sigma}$ is allowed to be complex to take into account circularly polarised laser beams as well. The Wigner function of this field is \footnote{This is derived by assuming the wave has some random initial phase which is averaged over. This averaging is equivalent to the mode truncation procedure which appears in the usual derivation \citep{kruer2019physics}, in which the coupling of the laser to plasma oscillations at $\omega \pm 2\omega_0$, $\omega \pm 3\omega_0$, etc. is ignored. The validity of this approximation was studied by \citet{machacek2001versatile}. There is an exactly equivalent issue in the DW-ZF problem which concerns the modulational instability of a single drift wave mode, and the derivation of the growth rate from the generalised zonostrophic instability dispersion relation. For details see \citet{parker2014thesis}. Other kinds of averaging which produce the same result are also possible \citep{dodin2022}.}: 
\begin{equation}
  	\overline{\bm{\mathsf{W}}}_{\boldsymbol{vv}}(\boldsymbol{k}) = \tfrac14 (2\pi)^3 v_{\text{os}}^2 \left[ \bm{\sigma} \bm{\sigma}^{\dagger} \delta(\boldsymbol{k} - \boldsymbol{k}_0) + \bm{\sigma}^* \bm{\sigma}^{\top} \delta(\boldsymbol{k} + \boldsymbol{k}_0) \right].
\end{equation}
Substituting this in (\ref{broadTPD}), and as usual ignoring the anti-Stokes term for being off-resonant, reduces to the familiar expression \citep[see][]{kruer2019physics}:
\begin{equation}\label{eq:monochromatic}
	D_e(\Omega,\boldsymbol{K}) D_e(\omega_0 - \Omega,\boldsymbol{K}-\boldsymbol{k}_0) = \frac{\omega_{\text{pe}}^2	v_{\text{os}}^2 |\boldsymbol{K} \bm{\cdot \bm{\sigma}}|^2}{4} \left( \frac{K^2 - |\boldsymbol{K}-\boldsymbol{k}_0|^2}{K |\boldsymbol{K}-\boldsymbol{k}_0|} \right)^{2},
\end{equation}
where one has used the approximation $\Omega \simeq \Omega_{-} \simeq \omega_{\text{pe}}$ in the numerator.

In the case of multiple monochromatic non-interfering laser beams, the Wigner function is (ignoring the anti-Stokes terms):
\begin{equation}
  	\overline{\bm{\mathsf{W}}}_{\boldsymbol{vv}}(\boldsymbol{k}) = \sum_{i} \tfrac14 (2\pi)^3 v_{\text{os},i}^2 \bm{\sigma}_i \bm{\sigma}_i^{\dagger} \delta(\boldsymbol{k} - \boldsymbol{k}_{0i}).
\end{equation}
The dispersion relation in this case becomes:
\begin{equation}
  	D_e(\Omega,\boldsymbol{K}) = \sum_{i} \left| \frac{|\boldsymbol{K}-\boldsymbol{k}_{0i}|^2 - K^2}{2 K |\boldsymbol{K}-\boldsymbol{k}_{0i}|} (\boldsymbol{K} \boldsymbol{\cdot} \bm{\sigma}_{i}) v_{\text{os},i} \right|^2 \frac{ \omega_{\text{pe}}^2}{D_e(\omega_0-\Omega,\boldsymbol{K}-\boldsymbol{k}_{0i})}.
\end{equation}
This agrees with the result of \citet{michel2013measured}.

\subsection{Single temporally incoherent beam}

Next, consider a single temporally incoherent laser beam. The Wigner function describing such a field is of the form (again ignoring the anti-Stokes term):
\begin{equation}\label{broadWigner}
	\overline{\bm{\mathsf{W}}}_{\boldsymbol{vv}}(\boldsymbol{k}) = \tfrac14 (2\pi)^3 v_{\text{os}}^2 \, \bm{\sigma} \bm{\sigma}^{\dagger} G \! \left( k_\parallel \right) \delta(\boldsymbol{k}_{\perp}),
\end{equation}
where $G(k_{\parallel})$ is the power spectrum centred around $k_0$ and normalised such that it integrates to unity; and also: $k_\parallel \doteq \boldsymbol{k} \boldsymbol{\cdot} \hat{\bm{\kappa}}$, $\boldsymbol{k}_\perp \doteq \boldsymbol{k} - k_\parallel \hat{\bm{\kappa}}$ with $\hat{\bm{\kappa}} \doteq \boldsymbol{k}_0/k_0$ being the unit vector in the direction of propagation. For $\Delta \omega / \omega_0 \ll 1$, the wavenumber spread $\Delta k$ is related to the frequency bandwidth through $\Delta \omega = v_{g0} \Delta k$ where $v_{g0} = \partial \omega_{\boldsymbol{k}} / \partial k |_{k_{0}}$ is the group velocity of electromagnetic waves at the peak of the power spectrum. Most commonly one considers broadband lasers with top-hat (flat), Lorentzian or Gaussian power spectra \citep{follett2019}. Upon substitution of (\ref{broadWigner}) in (\ref{broadTPD}), the dispersion relation has the form:
\begin{equation}\label{eq:disp}
  	D_e(\Omega,\boldsymbol{K}) = \frac{K_{\perp}^2v_{\text{os}}^2}{4} \mathscr{I}.
\end{equation}
Here, in the same way as before $\boldsymbol{K}_{\perp} \doteq \boldsymbol{K} - \boldsymbol{K}\boldsymbol{\cdot} \hat{\bm{\kappa}}$. The integral $\mathscr{I}$ is given by
\begin{equation}
	\mathscr{I} = \int_{C_\text{L}} \frac{\dif \omega_{k_{\parallel}}}{v_{g}(\omega_{k_{\parallel}})}  \frac{  \mathsf{C} \! \left (\omega_{k_{\parallel}} \right)  G \! \left(\omega_{k_{\parallel}} \right)  }{ \left( \omega_{k_{\parallel}} - \Omega \right)^2  - \omega_{e\boldsymbol{k}_-}^2 },
\end{equation}
in which the integration variables have been changed from $\boldsymbol{k}$ to $\omega_{\boldsymbol{k}}$  in order to deal with the pole in the integrand more easily; note that the integral should be taken along the Landau contour $C_{\text{L}}$. The functions $\mathsf{C}$ has been defined as $\mathsf{C} (k_{\parallel}) \doteq (K^2 \Omega_- - k_-^2 \Omega)^2/(K k_-)^2$. Note that $\mathsf{C}$ is a function of $\Omega$ and $\boldsymbol{K}$ as well, but those have been dropped for brevity.  Of course here by $k_-$ one means $k_- = |\boldsymbol{K} - k_{\parallel} \hat{\bm{\kappa}}|$. In order not to make the notation too cumbersome, functions such as $\mathsf{C}$, the spectrum $G$, or $\omega_{e\boldsymbol{k}_-}$ which are defined as functions of $k_{\parallel}$, one evaluates in terms of the frequency implicitly as follows: $\mathsf{C}(\omega_{k_\parallel}) \equiv \mathsf{C}[\mathsf{K}(\omega_{k_\parallel})]$, where $\mathsf{K}(\omega_{\boldsymbol{k}}) \doteq c^{-1} \sqrt{\omega_{\boldsymbol{k}}^2 - \omega_{\text{pe}}^2} = k$ allows one to go from the laser frequency to the laser wavenumber.

In the monochromatic case, the TPD growth rate maximises to a value of $\gamma_0 = \tfrac14 k_0 v_{\text{os}}$. This tells one that $\gamma_0/\omega_{\text{pe}}\ll 1$ since $\gamma_0/\omega_{\text{pe}} \sim k_0 v_{\text{os}}/\omega_{\text{pe}}  \sim v_{\text{os}}/c \ll 1$. The right hand side of (\ref{eq:disp}) is of order $\gamma_0^2 \ll \omega_{\text{pe}}^2$. Therefore it is possible to assume that the real part of the frequency satisfies the linear dispersion relation: $\Omega = \omega_{e\boldsymbol{K}} + \im \gamma$. With this the growth rate becomes:
\begin{equation}
  	\gamma = \frac{K_{\perp}^2v_{\text{os}}^2}{4} \frac{1}{2\omega_{e\boldsymbol{K}}} \text{Im}[\mathscr{I}],
\end{equation}
and the integral $\mathscr{I}$ is:
\begin{equation}\label{eq:integral}
	\mathscr{I} = \int_{C_\text{L}}  \frac{\dif \omega_{k_{\parallel}}}{v_{g}(\omega_{k_{\parallel}})} \,  \frac{  \mathsf{C} \! \left (\omega_{k_{\parallel}} \right)  G \! \left(\omega_{k_{\parallel}} \right)   }{ \left( \omega_{k_{\parallel}} - \omega_{e\boldsymbol{K}} - \im \gamma - \omega_{e\boldsymbol{k}_-} \right) \left( \omega_{k_{\parallel}} - \omega_{e\boldsymbol{K}} - \im \gamma + \omega_{e\boldsymbol{k}_-} \right) }. 
\end{equation}
The integrand of $\mathscr{I}$ has one pole of interest which occurs at a wavenumber $k_{\text{p}}(\boldsymbol{K})$ such that: $\omega_{k_{\text{p}}} - \omega_{e\boldsymbol{K}} - \omega_{e\boldsymbol{k}_{-}} = 0$. This is the frequency matching condition which for a given $k_{\text{p}}$ defines a circle in $\boldsymbol{K}$-space centred around $(K_{\parallel},K_{\perp}) = (k_0/2,0)$. In the limit of large bandwidth $\Delta \omega \gg \gamma$ one evaluates $\mathscr{I}$ using Plemelj's formula:
\begin{equation}
	\lim_{\varepsilon \to 0^+} \int_{-\infty}^{\infty} \frac{f(x)}{x-\zeta \mp \im \varepsilon} \, \dif x = \mathscr{P} \! \int_{-\infty}^{\infty} \frac{f(x)}{x-\zeta} \, \dif x \pm \im \pi f(\zeta),
\end{equation}
with $\mathscr{P} \! \int$ denoting the principal value integral and $\zeta, \varepsilon \in \mathbb{R}$. Physically this is due to the fact that for small growth rates, the resonance is very sharp and the instability is driven only by the part of the power spectrum which can perfectly satisfy the matching conditions. Evaluating the imaginary part at the pole:
\begin{equation}
  	\text{Im}[\mathscr{I}] = \pi \frac{\mathsf{C}(k_{\text{p}}) \, G(k_{\text{p}}) }{ v_{g}(k_{\text{p}}) \left( \omega_{k_{\text{p}}} - \omega_{e\boldsymbol{K}} + \omega_{e\boldsymbol{k}_-} \right)} \simeq \pi \frac{\mathsf{C}(k_{\text{p}}) \, G(k_{\text{p}}) }{ v_{g0} \,\omega_{0} }
\end{equation}
This gives the growth rate of the two-plasmon decay instability in a uniform plasma in the case of a single temporally incoherent laser beam:
\begin{equation}
  	\gamma_{\boldsymbol{K}} = \frac{\pi G(k_{\text{p}})}{v_{g0}} \,  \gamma^2_{0\boldsymbol{K}},
\end{equation}
where one uses the fact that: $\frac{K_{\perp}^2v_{\text{os}}^2}{8 \omega_{e\boldsymbol{K}} \omega_0} \mathsf{C}(k_{\text{p}},\boldsymbol{K}) \simeq \frac{K_{\perp}^2v_{\text{os}}^2}{16} \frac{(K^2 - |\boldsymbol{K} - k_{\text{p}} \hat{\bm{\kappa}}|^2)^2}{K^2 |\boldsymbol{K} - k_{\text{p}} \hat{\bm{\kappa}}|^2} = \gamma_{0\boldsymbol{K}}^2$, and $\gamma_{0\boldsymbol{K}}$ is the monochromatic growth rate. The reader is reminded here that the wavenumber at which the resonance occurs $k_{\text{p}}$ is a function of $\boldsymbol{K}$ and is determined from the frequency matching condition: $\omega_{k_{\text{p}}} - \omega_{e\boldsymbol{K}} - \omega_{e\boldsymbol{k}_{-}} = 0$.

Consider next the maximum value that $\gamma_{\boldsymbol{K}}$ attains over all wavenumbers $\boldsymbol{K}$. Evaluating that for the top-hat and Lorentzian power spectra: for the top hat $G(k_0) = 1/\Delta k$ and so the growth rate is: $\gamma_{\text{TH}} = \pi \gamma_0^2 / \Delta \omega$. For the Lorentzian $G(k) = \frac1\pi \frac{\Delta k/2}{(k-k_0)^2 + (\Delta k /2)^2}$ one has $G(k_0) = 2/\pi \Delta k$ and so: $\gamma_{\text{L}} = 2 \gamma_0^2/\Delta \omega$. Here $\gamma_0 = \tfrac14 k_0 v_{\text{os}}$ is the value at which $\gamma_{0\boldsymbol{K}}$ maximises. These results agree with the well known scaling of the growth rate as $ \gamma \sim \! \gamma_0^2/\Delta \omega$, but also include the dependence of the result on the laser spectral shape. It is interesting to note that while the coherence times for the two power spectra are (for the top hat, and Lorentzian respectively) $\tau_c = 2\pi/\Delta \omega, \, 2/\Delta \omega$ \citep[see also section \ref{sec:separate lines} for a definition of $\tau_{\text{c}}$]{goodman2015statistical,follett2019}, the growth rates differ by less than a factor of $\pi$. This shows that at least in the homogeneous case, it is not the coherence time per se which is the most relevant metric, but rather -- the fraction of laser energy which satisfies the resonance conditions. At first glance this is somewhat at odds with the results of \citet{follett2019}, which found that the coherence time provides a universal scaling for the absolute Raman and TPD thresholds; meaning that the thresholds were the same for fields with the same coherence times but different spectral shapes. More formally -- it appears that the peak growth rates of the instability scale with $G(k_0)$, and the absolute thresholds with $\int |G(k)|^2 \text{d} k$. The reason for this difference in behaviour is unknown but the problem of an absolute instability near the quarter critical in an inhomogeneous plasma is significantly more complicated and there is little reason to expect that it should be affected by bandwidth in the same way as the temporal growth rate in a homogeneous plasma, which has been calculated here.

\begin{figure}
	\includegraphics[width=\textwidth]{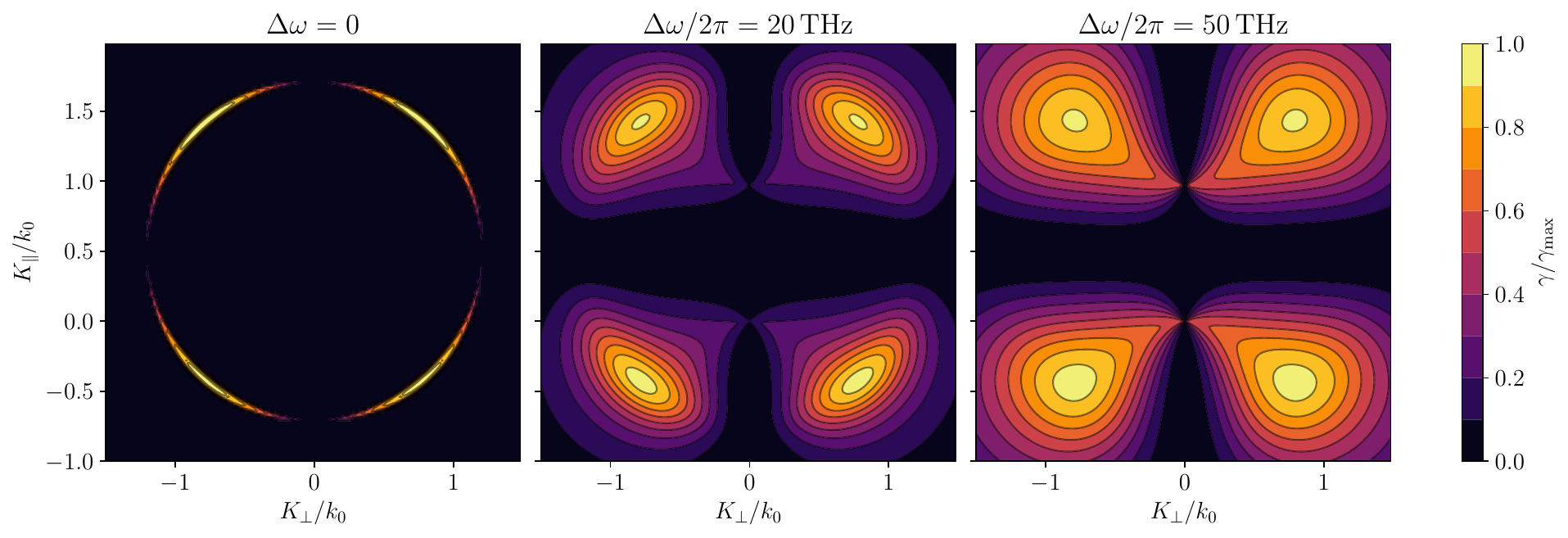}
	\caption{Effect of laser bandwidth on the range of unstable wavenumbers excited by the two-plasmon decay instability as driven by a single broadband laser beam. The laser has a Lorentzian power spectrum, an intensity of $I_{\text{L}} = 10^{15} \, \text{W}/\text{cm}^{2}$ and wavelength $\lambda = 350 \, \text{nm}$; the plasma conditions are: $T_e = 2 \, \text{keV}$ and $n_e = 0.23 n_\text{c}$ ($n_\text{c}$ being the critical density). The growth rates are normalised to the maximal growth rate corresponding to each configuration. The left panel represents the monochromatic case by solving (\ref{eq:monochromatic}) so it's normalised to $\gamma_{\text{max}} = \gamma_0 = \tfrac14 k_0 v_{\text{os}}$. The middle and right panels show the analytical broadband solution $\gamma_{\boldsymbol{K}}$ and are thus normalised to $\gamma_{\text{max}} = \gamma_{\text{L}} = 2 \gamma_0^2/\Delta \omega$.}
	\label{fig:kRange}
\end{figure}

Having explored how bandwidth affects the maximum value the growth rate attains, one now turns to how it modifies the spectrum of excited plasma waves. First in the monochromatic case, plasma waves are driven at wavenumbers near the intersection of the circles defined by the frequency matching condition, and the hyperbolas defined by $K_{\perp}^2 = K_{\parallel}(K_{\parallel} - k_0)$ at which the monochromatic growth rate $\gamma_{0\boldsymbol{K}}$ maximises. This is shown in the left panel in Figure \ref{fig:kRange}. In the middle and right panels the growth rate $\gamma_{\boldsymbol{K}}$ is presented, for a Lorentzian power spectrum of varying bandwidths. One sees that increasing the laser bandwidth significantly broadens the range of unstable wavenumbers excited by the two-plasmon instability. In the right panel virtually the entire TPD hyperbolas are observed since the ultra large laser bandwidth allows all relevant wavenumbers to satisfy the frequency matching conditions. What is most interesting is that even for relatively modest amounts of bandwidth such as $20 \, \text{THz}$ (middle panel) the broadening of the spectrum is substantial. Such effects may change the nature of the instability. For example, in an inhomogenous plasma, at lower densities the TPD modes are fairly high in $\boldsymbol{K}$ as illustrated by the left panel in figure \ref{fig:kRange}, and tend to be convective. Absolute modes occur near $|\boldsymbol{K}| \to 0$, which would typically be excited only in regions very close to the quarter critical \citep{yan2010linear}. One sees that laser bandwidth favours the absolute instability at lower densities too.

\begin{figure}
\centering
\includegraphics[width=0.8\textwidth]{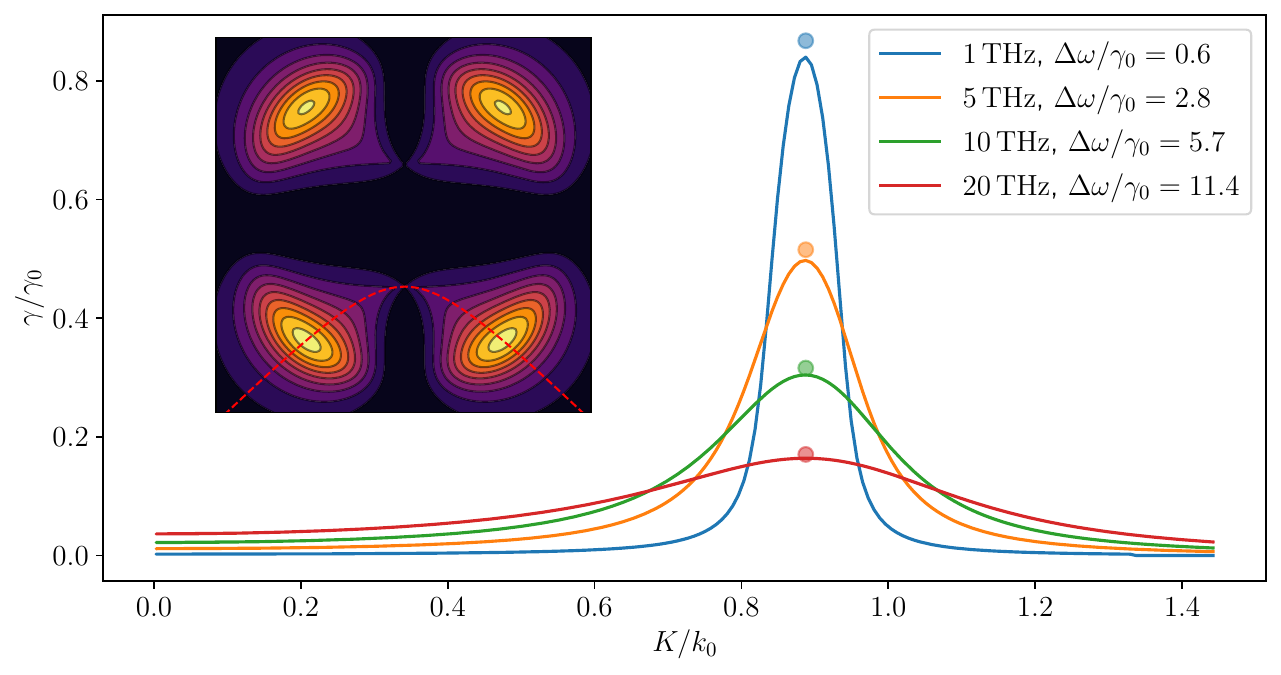}
\caption{Numerically evaluated growth rates $\gamma$ in the intermediate bandwidth regime calculated along the lower branch of the TPD hyperbolas defined by $K_{\perp}^2 = K_{\parallel}(K_{\parallel} - k_0)$. The inset shows the middle panel of figure \ref{fig:kRange} along with the aforementioned hyperbola as represented by the red dashed line. The plasma conditions and laser parameters are the same as the ones described in figure \ref{fig:kRange}. The growth rate is normalised to the monochromatic one $\gamma_0$, and the horizontal axis represents the magnitude of the wavenumber $K = \sqrt{K_\parallel^2 + K_{\perp}^2}$. The circles represent the estimate of the maximum growth rate as given by (\ref{eq:intermediate}).}
\label{fig:intermediate}
\end{figure}

The analytical results presented so far have relied upon the assumption that the bandwidth is large compared to the growth rate $\Delta \omega \gg \gamma_0$, which was used in the evaluation of the integral $\mathscr{I}$ in (\ref{eq:integral}). To explore the growth rates of the instability in the intermediate regime $\Delta \omega \sim \gamma_0$, the integral is evaluated numerically for finite $\gamma$. Figure \ref{fig:intermediate} shows the growth rate calculated along the lower branch of the TPD hyperbolas as a function of the magnitude of the wavenumber $K$. One sees that even at the fairly low amount of bandwidth of $5 \, \text{THz}$ ($\Delta \omega / \gamma_0 = 2.8$) and at intensity $I_{\text{L}} = 10^{15} \ \text{W}/\text{cm}^{2}$, the growth rate is half its monochromatic value. It is also approximately a third for $\Delta \omega/2\pi = 10 \, \text{THz}$ (corresponding to the intrinsic bandwidth of the argon-fluoride laser). As shown below, laser wavelength does not affect the peak values of the growth rate and therefore the results just quoted are unchanged by it. Decreasing the wavelength though does reduce the broadening effect in the plasma wave spectrum shown in figure \ref{fig:kRange}.

In addition, it is possible to calculate the peak of the growth rates analytically in this regime. This is done by ignoring the thermal correction in the dispersion functions and assuming the wavenumber $\boldsymbol{K}$ is along the TPD hyperbola, leading to an expression for the maximum growth rate of the form:
\begin{equation}
	\gamma_{\text{max}} = \gamma_0^2 \, \int \frac{2 \omega_{\text{pe}} \,G(\omega_{k_{\parallel}})}{(\omega_{k_{\parallel}} - 2 \omega_{\text{pe}} - \im \gamma_{\text{max}})(\omega_{k_{\parallel}} - \im \gamma_{\text{max}})} \, \dif \omega_{k_{\parallel}}.
\end{equation}
For a Lorentzian spectrum centred at $\omega_0 = 2 \omega_{\text{pe}}$ the integral can be evaluated exactly leading to:
\begin{equation}\label{eq:intermediate}
  	\gamma_{\text{max}} = \frac14 \left( \sqrt{16 \gamma_0^2 + \Delta \omega^2} - \Delta \omega \right).
\end{equation}
This reproduces the two relevant limits: for $\Delta \omega \to 0$, $\gamma_{\text{max}} \to \gamma_0$ and $\Delta \omega \to \infty$, $\gamma_{\text{max}} \to \gamma_{\text{L}} = 2\gamma_0^2/\Delta \omega$. Since $\gamma_0$ does not depend on laser wavelength, then $\gamma_{\text{max}}$ does not either. The circles in figure \ref{fig:intermediate} represent $\gamma_{\text{max}}$ as given by the result above, showing excellent agreement with the numerically calculated growth rate.

\subsection{Well separated discrete spectral lines} \label{sec:separate lines}

\begin{figure}
\centering
	\includegraphics[width=\textwidth]{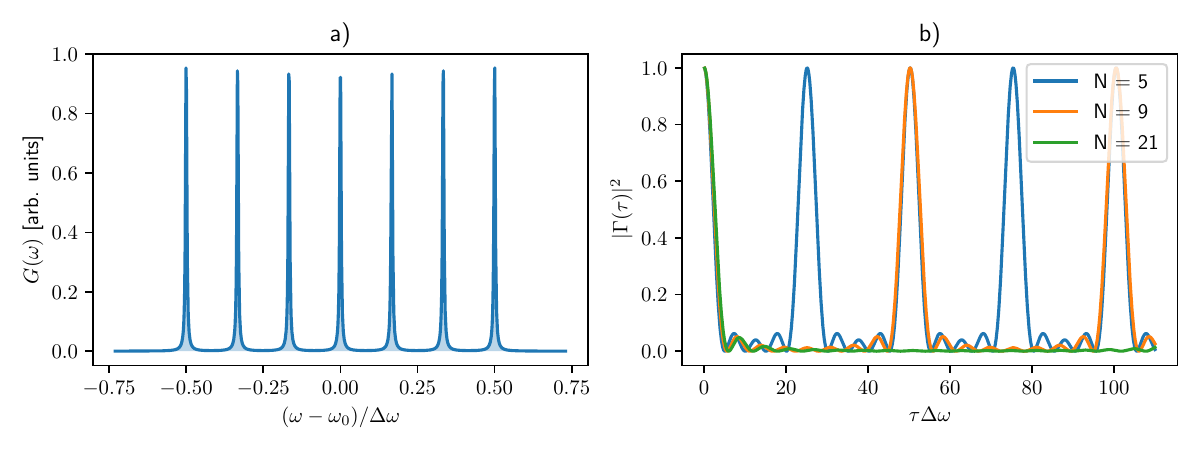}
	\caption{a) A power spectrum $G(\omega)$ consisting of $N=7$ well separated spectral lines, centred around $\omega_0$ and spread over a bandwidth of $\Delta\omega$. Their thickness is finite for the purposes of visualisation. b) The normalised correlation function $|\Gamma(\tau)|^2$ as a function of the time separation $\tau$ of a set of $N$ delta shaped spectral lines spread over a bandwidth $\Delta \omega$.}
	\label{fig:discrete}
\end{figure}

Another implication concerns the extent to which a power spectrum consisting of $N$ well separated sharp spectral lines spread over a bandwidth $\Delta \omega$, such as the one shown in figure \ref{fig:discrete}a), approximates the instability reduction properties of a continuous power spectrum \citep{bodner2023}. For sufficiently low number of spectral lines, such that only a single line is contained within the instability resonance: $\Delta \omega/(N-1) > \gamma_0$, following the generalised dispersion relation presented here, the instability growth rate will be $\gamma = \gamma_0/\sqrt{N}$. From the aforementioned inequality, this reduction will be somewhat lower than the expected $\pi \gamma_0^2/\Delta \omega$ in the continuous case, but still significant. For example for $N=10$, the reduction is similar to that of $10 \, \text{THz}$ continuous bandwidth (see figure \ref{fig:intermediate}). At first glance this reduction cannot be explained by the coherence properties of the power spectrum, since formally if each line is delta shaped, the coherence time is infinite.

To investigate this one follows \citet{goodman2015statistical}. Consider a power spectrum containing $N$ discrete delta-shaped spectral lines, uniformly spread over a bandwidth of $\Delta \omega$, and centred around $\omega_0$ (figure \ref{fig:discrete}a):
\begin{equation}
	G(\omega) = \frac{1}{N} \sum_{j=0}^{N-1} \delta\!\left( \omega - \omega_{0} - \frac{\Delta \omega}{2} + \frac{\Delta \omega}{N-1} j \right).
\end{equation}
This is the power spectrum of the complex analytic signal $Y(t)$ where the actual signal is $y(t) = \text{Re} \, Y(t)$. By the Wiener-Khinchin theorem, the correlation function $C(\tau) = \overline{Y^*(t) Y(t + \tau)}$ is given by the (inverse) Fourier transform of $G(\omega)$:
\begin{equation}
	C(\tau) = \frac{1}{2\pi} \int G(\omega) e^{- \im \omega t} \dif t.
\end{equation}
Defining the normalised correlation function $\Gamma(\tau) = C(\tau)/C(0)$, the coherence time for a random process is given by
\begin{equation}
	\tau_c = \int_{-\infty}^{\infty} |\Gamma(\tau)|^2 \dif \tau.
\end{equation}
It can be shown that the correlation function for the discrete power spectrum $G(\omega)$ is:
\begin{equation}
	|\Gamma(\tau)|^2 = \frac{1}{N^2} \csc^2 \left( \frac{\Delta \omega \tau}{2(N-1)} \right) \sin^2 \left( \frac{N \Delta \omega \tau}{2(N-1)} \right). 
\end{equation}
Formally, the integral of this will always diverge for any finite $N$. In the limit of $N \to \infty$ one recovers the normalised correlation function for a continuous top hat spectrum which has a finite coherence time:
\begin{equation}
	|\Gamma_{\text{TH}}(\tau)|^{2} = \text{sinc}^{2} (\Delta \omega \tau/2).
\end{equation} 
Figure \ref{fig:discrete}b) shows what $|\Gamma(\tau)|^2$ looks like for different number of spectral lines. One sees that for small $\tau$ the correlation functions closely resemble the sinc profile expected in the continuous case. In contrast to it, the discrete case exhibits correlations for larger $\tau$, and the timescale over which these correlations occur increases with the number of spectral lines $N$. The $k^{\text{th}}$ peak occurs at time delay $T_k = 2(N-1) \pi k / \Delta \omega$. If $T_1$, the time at which the first peak occurs, is much larger than the typical time scale of interest (such as $\gamma_0^{-1}$) then the coherence properties of the continuous spectrum are well approximated by the discrete one. Using $T_1 \gg \gamma_0^{-1}$ translates to a condition on the number of spectral lines:
\begin{equation}
  	N \gg \frac{1}{2\pi} \frac{\Delta \omega}{\gamma_0}.
\end{equation}
For a laser intensity of $10^{15} \, \text{W} \, \text{cm}^{-2}$ at a wavelength of $350 \, \text{nm}$ and bandwidth of $\Delta \omega/\omega_0 = 1 \, \%$, the right hand side is $\simeq 1$. This suggests that even a low number of discrete spectral lines such as $N = 10$ might be a good approximation to the coherence properties of the continuous spectrum, as far as their effect on the instability is concerned. 

\section{Summary}
\label{sec:summary}

In this paper, a novel statistical theory of the broadband two-plasmon decay instability has been presented. It has been used to derive a dispersion relation in uniform plasma that is valid under laser electromagnetic fields with arbitrary power spectra. To achieve this, recent developments in inhomogeneous turbulence have been applied in order to derive the statistical closure required to treat the problem. The new dispersion relation is then applied to the study of the instability in a few cases of interest. For a single temporally incoherent laser beam, in the limit of large bandwidth $\Delta \omega \gg \gamma_0$, the well known reduction of the peak of the growth rate by a factor of approximately $\gamma_0/\Delta \omega$ is confirmed, but it's also shown that the exact amount depends on the shape of the power spectrum. Namely for a Lorentzian power spectrum the peak growth rate is $\gamma_{\text{L}} = 2 \gamma_0^2/\Delta \omega$, and for a flat (top-hat) spectrum it's $\gamma_{\text{TH}} = \pi \gamma_0^2/\Delta \omega$. This difference is less than what the difference in the coherence times of the two laser beams would imply. In addition, it has been shown that the laser bandwidth significantly broadens the range of wavenumbers excited by TPD. This implies that bandwidth favours the absolute instability in regions further away from the quarter critical density by allowing the $|\boldsymbol{K}| \to 0$ modes to be excited there too. The intermediate bandwidth regime is explored numerically and it is shown that the growth rate is reduced to half its value for laser intensities of $10^{15} \, \text{W}/\text{cm}^{2}$ and relatively modest bandwidths of $5 \, \text{THz}$. A power spectrum consisting of $N$ discrete lines spread over a bandwidth of $\Delta \omega$ is also studied. It is shown that for a low number of spectral lines, the reduction in the peak growth rate is $1/\sqrt{N}$, which is somewhat smaller than in the continuous case but still significant. 

There some natural questions which arise as a consequence of this work. Firstly, can one model transversely incoherent laser fields such as those produced by a random phase plate (RPP), or a combination of transverse and temporal incoherence such as when a phase plate is combined with smoothing by specral dispersion (SSD), as well as in the induced spatial coherence scheme (ISI)? Formally transverse incoherence can be straightforwardly incorporated by simply allowing the Wigner function (equivalently, the laser power spectrum) to have some finite angular spread in wavenumber. But since two laser fields which can have the same power spectrum, can nonetheless still be very different, care should be taken when applying the generalised dispersion relation. The RPP field is such that it creates a speckle pattern at the focus which may cause the instability to be strongly localised within speckles with many times the average intensity \citep{follett2022independent}. Such intermittent behaviour may break the Gaussianity assumption made in this work. Adding temporal incoherence by SSD or by using naturally broad-bandwidth lasers such as the excimer ones used in the ISI scheme, causes the speckle pattern to move which may help smooth out this behaviour. 

Secondly, the presence of a density gradient is primarily responsible for determining whether the dominant instability at the quarter critical is the two plasmon decay or Raman scattering, with longer, ignition-relevant scale lengths favouring the Raman instability \citep{rosenberg2018}. Therefore extending the present analysis to study the effects of laser bandwidth on the absolute instability in the presence of a density gradient is an important problem. The approach of \citet{liu1976} and \cite{simon1983inhomogeneous} for a plane wave laser field uses some rather specific mathematical manipulations to calculate the absolute threshold, for which it isn't immediately clear how they might be carried out within the framework presented in this paper. So this remains an interesting open problem.

\section*{Acknowledgements}

The authors acknowledge useful discussions with Dr. Russell Follett and Dr. John Palastro (Laboratory for Laser Energetics, Univ. Rochester), Dr. Steven Bodner (Naval Research Laboratory - Emeritus) and Dr. Archie Bott (Oxford), as well as the two anonymous referees for their valuable input.

\section*{Funding}

R.T.R. was supported by the Saven European Scholarship and UKRI-EPSRC. R.A. was supported by the Oxford-ShanghaiTech collaboration agreement. The research was conducted under UKRI-grants EP/X035336/1 and ST/V001655/1.

\section*{Declaration of interests}

The authors report no conflict of interest.

%%%% APENDICIES %%%%

\appendix

\section{The Weyl symbol calculus}
\label{appWeyl}

Here a very brief introduction to the Weyl symbol calculus is provided. Although quite concise, this should give the reader most of the necessary background to understand the results in this paper. For details and proofs see \citet{mcdonald1988,tracy2014ray, dodin2022}, and references therein.

The symbol calculus is best presented in terms of abstract bra-kets and abstract operators acting on them. Therefore for any fields which might be of interest -- the plasma density perturbation $n$, the velocity potential $\psi$, or the quiver velocity $\boldsymbol{v}_{\text{os}}$, etc. -- one considers their abstract bra-ket representations. So for example, the position representation of the density perturbation is $n(\boldsymbol{x}) = \braket{\boldsymbol{x}}{n}$; it's Fourier transform, i.e. its wavenumber representation, is $\mathring{n}(\boldsymbol{k}) = \braket{\boldsymbol{k}}{n}$. 

One starts by defining the Wigner transform $\mathscr{W}$ of an abstract operator $\hat{A}$ as follows:
\begin{equation}
  	\mathsf{A}(\boldsymbol{x},\boldsymbol{k}) \doteq \mathscr{W}[\hat{A}] \doteq \int \dif \boldsymbol{s} \, e^{- \im \boldsymbol{k} \boldsymbol{\cdot} \boldsymbol{s}} \bra{\boldsymbol{x} + \boldsymbol{s}/2} \hat{A} \ket{\boldsymbol{x} - \boldsymbol{s}/2},
\end{equation}
where $\mathsf{A}(\boldsymbol{x},\boldsymbol{k})$ is called the \textit{Weyl symbol} of the operator $\hat{A}$. There is a close connection between the Wigner transform and correlation functions. The Wigner function of some field $\psi(\boldsymbol{x})$ is defined as:
\begin{equation}
  	\mathsf{W}_{\psi \psi}(\boldsymbol{x},\boldsymbol{k}) \doteq \int \dif \boldsymbol{s} \, e^{- \im \boldsymbol{k} \boldsymbol{\cdot} \boldsymbol{s}} \, \psi \! \left( \boldsymbol{x} + \boldsymbol{s}/2 \right) \psi^\dagger \! \left( \boldsymbol{x} - \boldsymbol{s}/2 \right),
\end{equation}
where from the definition, it is apparent that it can be thought of as the Fourier transform of the two-point correlation function of $\psi$ (modulo some averaging procedure). The Wigner function is the symbol of the density operator $\hat{\rho} \doteq \ket{\psi} \bra{\psi} \,$:
\begin{equation}
  	\mathsf{W}_{\psi \psi}(\boldsymbol{x},\boldsymbol{k}) = \mathscr{W}\! \left[ \hat{\rho} \right].
\end{equation}
Analogously, one describes correlations between two different fields $\psi$ and $\phi$ by considering the Wigner transform of $\ket{\psi} \bra{\phi}$ :
\begin{equation}
	\mathsf{W}_{\psi \phi}(\boldsymbol{x},\boldsymbol{k}) = \mathscr{W}\! \left[ \ \ket{\psi} \bra{\phi} \ \right] = \int \dif \boldsymbol{s} \, e^{- \im \boldsymbol{k} \boldsymbol{\cdot} \boldsymbol{s}} \, \psi \! \left( \boldsymbol{x} + \boldsymbol{s}/2 \right) \phi^\dagger \! \left( \boldsymbol{x} - \boldsymbol{s}/2 \right).
\end{equation}

The Wigner transform possesses some nice properties:
\begin{equation}
  	\mathscr{W}[\mathbb{1}] = 1, \quad \mathscr{W}[\hat{x}_j^n] = x_j^n, \quad \mathscr{W}[\hat{k}_j^n] = k_j^n,
\end{equation}
for any integer $n$, where $\mathbb{1}$ is the identity operator, $\hat{\boldsymbol{x}}$ is the abstract position operator with $\hat{x}_j$ being its $j^{\text{th}}$ component; analogously for the wavenumber operator $\hat{\boldsymbol{k}}$, which is the abstract representation of $- \im \partial_{\boldsymbol{x}}$\footnote{This shows how similarly to Fourier transforms, when taking the Wigner transform of an equation, derivatives can be simply substituted for $\im \boldsymbol{k}$.}. As a consequence, any reasonably behaved function $F$ of those operators satisfies:
\begin{equation}
  	\mathscr{W}[F(\hat{\boldsymbol{x}})] = F(\boldsymbol{x}), \quad \quad \mathscr{W}[F(\hat{\boldsymbol{k}})] = F(\boldsymbol{k}).
\end{equation}
This property makes it particularly easy to go from equations in the position representation as they are most commonly given, to their abstract representation which is suitable for applying the Weyl symbol calculus. In the present work, this step does not even need be made explicit.

The symbol of a product of operators $\mathscr{W}[\hat{A} \hat{B}]$ is related to the individual symbols $\mathscr{W}[\hat{A}] = \mathsf{A}(\boldsymbol{x},\boldsymbol{k})$ and $\mathscr{W}[\hat{B}] = \mathsf{B}(\boldsymbol{x},\boldsymbol{k})$ through the \textit{Moyal star product} $\star$ as follows:
\begin{equation}
  	\mathscr{W}[\hat{A} \hat{B}] = \mathsf{A}(\boldsymbol{x},\boldsymbol{k}) \star \mathsf{B}(\boldsymbol{x},\boldsymbol{k}),
\end{equation}
where the star product between two symbols is defined as
\begin{equation}
  	\mathsf{A}(\boldsymbol{x},\boldsymbol{k}) \star \mathsf{B}(\boldsymbol{x},\boldsymbol{k}) \doteq  \mathsf{A}(\boldsymbol{x},\boldsymbol{k}) e^{\im \hat{\mathcal{P}}/2}  \mathsf{B}(\boldsymbol{x},\boldsymbol{k}), \quad \quad \hat{\mathcal{P}} = \oleft{\partial_{\boldsymbol{x}}} \boldsymbol{\cdot} \oright{\partial_{\boldsymbol{k}}} - \oleft{\partial_{\boldsymbol{k}}} \boldsymbol{\cdot} \oright{\partial_{\boldsymbol{x}}},
\end{equation}
with $\hat{\mathcal{P}}$ being the Poisson bracket, and the arrows over the derivatives denoting the direction in which they act; $\mathsf{A} \hat{\mathcal{P}} \mathsf{B}$ represents the Poisson bracket between $\mathsf{A}$ and $\mathsf{B}$. The symbol product and the Wigner transform are related to each other like the convolution and the Fourier transform. 

Frequently symbol products exist where one of the symbols is only a function of space, or only a function of wavenumber. In such cases, the symbol product simplifies greatly when Fourier transformed. Denoting the Fourier transform of $\psi(\boldsymbol{x})$ as $\mathring{\psi}(\boldsymbol{k}) \doteq \mathscr{F}_{\boldsymbol{K}} \! \left[ \psi(\boldsymbol{x}) \right] = \int \! \dif \boldsymbol{x}\, \psi(\boldsymbol{x}) e^{ - \im \boldsymbol{K} \boldsymbol{\cdot} \boldsymbol{x}}$ one has:
\begin{equation}\label{symbFT}
	\begin{aligned}
		\mathscr{F}_{\boldsymbol{K}} \! \left[ F(\boldsymbol{k}) \star W(\boldsymbol{x},\boldsymbol{k}) \right] &= F \! \left( \boldsymbol{k} - \tfrac12 \boldsymbol{K} \right) \mathring{W}(\boldsymbol{K},\boldsymbol{k}), \\
		\mathscr{F}_{\boldsymbol{K}} \! \left[  W(\boldsymbol{x},\boldsymbol{k}) \star F(\boldsymbol{k}) \right] &= F \! \left( \boldsymbol{k} + \tfrac12 \boldsymbol{K} \right) \mathring{W}(\boldsymbol{K},\boldsymbol{k}), \\
		\mathscr{F}_{\boldsymbol{K}} \! \left[ G(\boldsymbol{x}) \star W(\boldsymbol{k}) \right] &= \mathring{G}(\boldsymbol{K}) W \! \left( \boldsymbol{k} + \tfrac12 \boldsymbol{K} \right),
	\end{aligned}
\end{equation}
for some generic functions $F$, $G$ and $W$. The proof of the first one is as follows:
\begin{equation*}
	\begin{aligned}
		\mathscr{F}_{\boldsymbol{K}} \! \left[ F(\boldsymbol{k}) \star W(\boldsymbol{x},\boldsymbol{k}) \right] &= \mathscr{F}_{\boldsymbol{K}} \! \left[ F(\boldsymbol{k}) \exp \left( \frac{\im \oleft{\partial}_{\boldsymbol{x}} \bm{\cdot} \oright{\partial}_{\boldsymbol{k}} - \im \oleft{\partial}_{\boldsymbol{k}} \bm{\cdot} \oright{\partial}_{\boldsymbol{x}} }{2} \right) W(\boldsymbol{x},\boldsymbol{k}) \right] \\
		&= F(\boldsymbol{k}) \mathscr{F}_{\boldsymbol{K}} \! \left[ \exp \left( - \tfrac12 \im \oleft{\partial}_{\boldsymbol{k}} \bm{\cdot} \oright{\partial}_{\boldsymbol{x}} \right) W(\boldsymbol{x},\boldsymbol{k}) \right] \\
		&= F(\boldsymbol{k}) \exp \left( - \tfrac12 \oleft{\partial}_{\boldsymbol{k}} \bm{\cdot} \boldsymbol{K} \right) \mathring{W}(\boldsymbol{K},\boldsymbol{k}) \\
		&= F \! \left( \boldsymbol{k} - \tfrac12 \boldsymbol{K} \right) \mathring{W}(\boldsymbol{K},\boldsymbol{k}).
	\end{aligned}
\end{equation*}
The rest follow analogously.

For some field $\ket{\psi}$ governed by a Schr\"odinger-like equation $\im \partial_t \ket{\psi} = \hat{H} \ket{\psi}$ the density operator satisfies the von-Neumann equation:
\begin{equation}
  	\im \partial_t \hat{\rho} = \hat{H} \hat{\rho} - \hat{\rho} \hat{H}^{\dagger}.
\end{equation}
Taking the Wigner transform of the above gives the so called \textit{Wigner-Moyal equation} which governs the Wigner function of the field:
\begin{equation}
  	\im \partial_t \mathsf{W}_{\psi \psi} = \mathsf{H} \star \mathsf{W}_{\psi \psi} - \mathsf{W}_{\psi \psi} \star \mathsf{H}^{\dagger} \tag{WME},
\end{equation}
with $\mathsf{H} \doteq \mathscr{W}[\hat{H}]$ being the symbol of the Hamiltonian operator $\hat{H}$. This is what allows us to very efficiently go from the equations of motion (EOM) of some field $\psi$, to the equation governing the correlation function of that field:
\begin{equation*}
	\text{EOM of $\psi$} \quad \xrightarrow{\text{Weyl symbol calculus}} \quad \text{EOM of the correlation function of $\psi$}.
\end{equation*}

All of the above properties are also satisfied for vector fields $\bm{\Psi} = (\psi_1, \psi_2, \dots)$ and matricies of operators $\hat{\bm{A}}$ as long as the usual matrix multiplication rules are respected. For example
\begin{equation}
  \left( \bm{\mathsf{A}} \star \bm{\mathsf{B}} \right)_{ij} =  \sum_k (\bm{\mathsf{A}})_{ik} \star (\bm{\mathsf{B}})_{kj},
\end{equation}
where $\bm{\mathsf{A}}$ and $\bm{\mathsf{B}}$ are arbitrary matrix symbols. Matrix multiplication rules will be implied between symbols from now on.

\section{The driving terms}
\label{appDriving}

Here a calculation identical to the one in \citet{tsiolis2020structure} is reproduced which gives the form of the TPD driving terms. Consider the following
\begin{equation}
  	\begin{aligned}
  		\bm{v}_{\text{os}} \bm{\cdot} \bm{\grad} \psi &= \sum_j \im  \bra{\boldsymbol{x}} \hat{k}_j \ket{\psi} \braket{\boldsymbol{x}}{v_{\text{os};j}} \\
  		&= \sum_j\im \bra{\boldsymbol{x}} \hat{k}_{j} \ket{\psi} \braket{v_{\text{os};j}}{\boldsymbol{x}} \\
  		&= \im \bra{\boldsymbol{x}} \hat{\boldsymbol{k}} \boldsymbol{\cdot} \hat{\bm{W}}_{\psi \boldsymbol{v}} \ket{\boldsymbol{x}}.
  	\end{aligned}
\end{equation}
here $\hat{\bm{W}}_{\psi \bm{v}} \doteq \ket{\psi}\bra{\bm{v}_{\text{os}}} \,$. So one sees that the expression $\bm{v}_{\text{os}} \bm{\cdot} \bm{\grad} \psi$ can be thought of as the matrix elements of the operator $\im \hat{\boldsymbol{k}} \boldsymbol{\cdot} \hat{\bm{W}}_{\psi \boldsymbol{v}}$. For a general operator $\hat{A}$, it can be shown that these matrix elements can be related to the Weyl symbol $\mathsf{A}(\boldsymbol{x},\boldsymbol{k})$ of the operator as follows \citep{dodin2022}:
\begin{equation}
	\bra{\boldsymbol{x}} \hat{A} \ket{\boldsymbol{x}} = \frac{1}{(2\pi)^3} \int \dif \boldsymbol{k} \, \mathsf{A}(\boldsymbol{x},\boldsymbol{k}).
\end{equation}
Hence using the Moyal star product, one writes
\begin{equation}
	\boldsymbol{v}_{\text{os}} \boldsymbol{\cdot} \grad \psi = \im \int \frac{\dif \boldsymbol{k}}{(2\pi)^3} \, \boldsymbol{k}^{\top} \star \bm{\mathsf{W}}_{\psi \boldsymbol{v}}^\top,
\end{equation}
since
\begin{equation}
  	\mathscr{W} \left[ \hat{\boldsymbol{k}} \boldsymbol{\cdot} \hat{\bm{W}}_{\psi \boldsymbol{v}} \right] = \boldsymbol{k}^{\top} \star \bm{\mathsf{W}}_{\psi \boldsymbol{v}}^\top.
\end{equation}
Note that the transpose is needed in order for the dot product to be taken into account in the symbol product since $\boldsymbol{k}$ is a column vector, and $\bm{\mathsf{W}}_{\psi \boldsymbol{v}}$ is a row vector.

\section{Proof of the relationship between $\boldsymbol{\mathsf{B}}_{\boldsymbol{vv}} \doteq \bm{\mathsf{W}}_{(\bm{\phi} + \bm{\chi)}(\bm{\phi} - \bm{\chi)}}$ and $\bm{\mathsf{W}}_{\boldsymbol{v}\boldsymbol{v}} = \bm{\mathsf{W}}_{(\bm{\phi} + \bm{\chi)}(\bm{\phi} + \bm{\chi)}} $}
\label{appProofPump}

The solution of the Klein-Gordon equation for the vector potential in a uniform plasma is of the form
\begin{equation}
	\boldsymbol{A}(t,\boldsymbol{x}) = \int \frac{\dif \boldsymbol{k}}{(2\pi)^3} \, \mathring{\boldsymbol{A}}(\boldsymbol{k}) e^{ - \im \omega_{\boldsymbol{k}} t + \im \boldsymbol{k} \boldsymbol{\cdot} \boldsymbol{x} },
\end{equation}
with $\omega_{\boldsymbol{k}} = \sqrt{\omega_{\text{pe}}^{2} +  k^2 c^2}$ being the photon frequency. To calculate $\boldsymbol{\mathsf{B}}_{\boldsymbol{vv}}$ one needs $\partial_t \boldsymbol{v}_{\text{os}}$ which is given by
\begin{equation}
	\partial_t \boldsymbol{v}_{\text{os}} =  \int \frac{\dif \boldsymbol{k}}{(2\pi)^3} \, \left( - \im \omega_{\boldsymbol{k}} \right) \mathring{\boldsymbol{v}}_{\text{os}}(\boldsymbol{k})  e^{ - \im \omega_{\boldsymbol{k}} t + \im \boldsymbol{k} \boldsymbol{\cdot} \boldsymbol{x} }.
\end{equation}
Let $\mathring{\boldsymbol{V}}$ denote the Fourier transform of $\partial_t \boldsymbol{v}_{\text{os}}$:
\begin{equation}
	\mathring{\boldsymbol{V}}(\boldsymbol{k}) \doteq \mathscr{F}_{\boldsymbol{k}} \left[ \partial_t \boldsymbol{v}_{\text{os}} \right] = - \im \omega_{\boldsymbol{k}} \mathring{\boldsymbol{v}}_{\text{os}}(\boldsymbol{k}).
\end{equation}
It's straightforward to show that the Wigner function can be expressed in terms of the Fourier transforms of the fields as follows:
\begin{equation}
  	\mathsf{W}_{\psi \phi}(\boldsymbol{x},\boldsymbol{k}) = \int \dif \boldsymbol{k}' \, e^{\im \boldsymbol{k}' \boldsymbol{\cdot} \boldsymbol{x}} \, \mathring{\psi} \! \left( \boldsymbol{k} + \boldsymbol{k}'/2 \right) \mathring{\phi}^\dagger \! \left( \boldsymbol{k} - \boldsymbol{k}'/2 \right).
\end{equation}
Therefore:
\begin{equation}
\begin{aligned}
		\boldsymbol{\mathsf{B}}_{\boldsymbol{vv}} &= \im \omega_{\text{pe}}^{-1} \int \dif \boldsymbol{k}' \, e^{\im \boldsymbol{k}' \bm{\cdot} \boldsymbol{x}} \, \mathring{\boldsymbol{v}}_{\text{os}}(\boldsymbol{k} + \boldsymbol{k}'/2) \mathring{\boldsymbol{V}}^{\dagger}(\boldsymbol{k}-\boldsymbol{k}'/2)  \\
	&= - \omega_{\text{pe}}^{-1} \int \dif \boldsymbol{k}' \, e^{\im \boldsymbol{k}' \bm{\cdot} \boldsymbol{x}} \, \omega_{\boldsymbol{k}-\boldsymbol{k}'/2} \, \mathring{\boldsymbol{v}}_{\text{os}}(\boldsymbol{k} + \boldsymbol{k}'/2) \mathring{\boldsymbol{v}}_{\text{os}}^{\dagger}(\boldsymbol{k}-\boldsymbol{k}'/2).
\end{aligned}
\end{equation}
The equivalent expression for $\bm{\mathsf{W}}_{\boldsymbol{vv}}$ is
\begin{equation}
	\bm{\mathsf{W}}_{\boldsymbol{vv}} = \int \dif \boldsymbol{k}' \, e^{\im \boldsymbol{k}' \bm{\cdot} \boldsymbol{x}} \, \mathring{\boldsymbol{v}}_{\text{os}}(\boldsymbol{k} + \boldsymbol{k}'/2) \mathring{\boldsymbol{v}}_{\text{os}}^{\dagger}(\boldsymbol{k}-\boldsymbol{k}'/2).
\end{equation}
For spatially homogeneous fields, upon statistical averaging these wavenumber correlators will turn out to be deltas. To see that consider
\begin{equation}
	\begin{aligned}
		\overline{\boldsymbol{a}(\boldsymbol{x}+\boldsymbol{s}/2)  \boldsymbol{a}^\dagger(\boldsymbol{x}-\boldsymbol{s}/2)} &= \int \dif \boldsymbol{k}_1 \dif \boldsymbol{k}_2 \,  \overline{\mathring{\boldsymbol{a}} (\boldsymbol{k}_2) \mathring{\boldsymbol{a}}^\dagger(\boldsymbol{k}_1)} e^{\im \boldsymbol{k}_1 \bm{\cdot} (\boldsymbol{x}-\boldsymbol{s}/2)} e^{-\im \boldsymbol{k}_2 \bm{\cdot} (\boldsymbol{x}+\boldsymbol{s}/2)} \\
		&= \int \dif \boldsymbol{k}_1 \dif \boldsymbol{k}_2 \,  \overline{\mathring{\boldsymbol{a}} (\boldsymbol{k}_2) \mathring{\boldsymbol{a}}^\dagger(\boldsymbol{k}_1)}  e^{\im (\boldsymbol{k}_1 - \boldsymbol{k}_2) \bm{\cdot} \boldsymbol{x}} e^{- \im (\boldsymbol{k}_1 + \boldsymbol{k}_2)\bm{\cdot} \boldsymbol{s}/2} \\
		&= \int \dif \bar{\boldsymbol{k}} \dif \boldsymbol{k}' \,\overline{\mathring{\boldsymbol{a}} (\bar{\boldsymbol{k}} - \boldsymbol{k}'/2) \mathring{\boldsymbol{a}}^\dagger(\bar{\boldsymbol{k}} + \boldsymbol{k}'/2)} e^{\im \boldsymbol{k}' \bm{\cdot} \boldsymbol{x}} e ^{-\im \bar{\boldsymbol{k}}\bm{\cdot}\boldsymbol{s}}.
 	\end{aligned}
\end{equation}
For this to be independent of $\boldsymbol{x}$ one needs 
\begin{equation}
	\overline{\mathring{\boldsymbol{a}} (\bar{\boldsymbol{k}} - \boldsymbol{k}'/2) \mathring{\boldsymbol{a}}^\dagger(\bar{\boldsymbol{k}} + \boldsymbol{k}'/2)} \propto \delta(\boldsymbol{k}').
\end{equation}
Hence after statistical averaging and integrating out the delta, one gets the required relationship:
\begin{equation}
	\overline{\boldsymbol{\mathsf{B}}}_{\boldsymbol{vv}}(\boldsymbol{k}) = - \omega_{\text{pe}}^{-1} \omega_{\boldsymbol{k}} \overline{\bm{\mathsf{W}}}_{\boldsymbol{vv}}(\boldsymbol{k}).
\end{equation}

%\nocite{*}
\bibliographystyle{jpp}

\bibliography{references.bib}

\end{document}